\documentclass[twocolumn,showpacs,preprintnumbers,amsmath,amssymb]{revtex4}
\usepackage{dsfont}
\usepackage{cases}
\usepackage{graphicx}% Include figure files
\usepackage{dcolumn}% Align table columns on decimal point
\usepackage{bm}% bold math
\usepackage{amsmath}

\UseRawInputEncoding
\begin{document}

\title{Entanglement Dynamics of Two V-type Atoms with Dipole-Dipole Interaction in Dissipative Cavity}
\author{Jia Wang}
\author{Dan Long}
\author{Qilin Wang}
\author{ Hong-Mei Zou}%
\email{zhmzc1997@hunnu.edu.cn}
\author{Chenya Liu}
\author{Qianqian Ma}%
 \affiliation{Synergetic Innovation Center for Quantum Effects and Application, Key Laboratory of Low-dimensional Quantum Structures and Quantum Control of Ministry of Education, School of Physics and Electronics, Hunan Normal University, Changsha, 410081, People's Republic of China.}%

\date{\today}% It is always \today, today,
\begin{abstract}
In this work, we study a coupled system of two V-type atoms with dipole-dipole interaction in a dissipative single-mode cavity, which couples with an external environment. We obtain the analytical solution of this model by solving the time dependent Schrodinger equation after we diagonalize Hamiltonian of dissipative cavity by introducing a set of new creation and annihilation operators according to Fano theorem. We also detailedly discuss the influences of cavity-environment coupling, spontaneously generated interference (SGI) parameter and dipole-dipole interaction between two atoms on entanglement dynamics under different initial states. The results show that the SGI parameter has different effects on entanglement dynamics under different initial states. Namely, the SGI parameter will increase the decay rate of the initially maximal entangled state and reduce that of the initially  partial entangled state. For the initially product state, the larger SGI parameter corresponds to the more entanglement generated. The entanglement monotonically decreases under the weak cavity-environment coupling, while the oscillation of entanglement will occur under the strong cavity-environment coupling. The larger the dipole-dipole interaction is, the slower the entanglement decays and the more the entanglement will be generated. So the dipole-dipole interaction can not only protect and generate entanglement very effectively, but also enhance the regulation effect of the SGI parameter on entanglement.
\end{abstract}
\pacs{03.65.Yz, 03.67.Lx, 42.50.-p, 42.50.Pq.}

\maketitle

\section{Introduction}

Quantum entanglement is a pure quantum property different from classical physics and provides a powerful physical resource for quantum information. In the past three decades, there are many new advances in quantum information based on quantum entanglement. For example, some schemes of quantum teleportation have been proposed theoretically and experimentally \cite{Pirandola S,Bouwmeester D,Yan Z H,Cacciapuoti A S,Llewellyn D,Lipka-Bartosik P}. In quantum key distribution, since Bennett and Brassard proposed the BB84 protocol \cite{C.H. Bennett}, many new methods which can solve the quantum key distribution problem have been proposed successively \cite{Scarani V,Renner R,Manzalini A,Fitzke E,Neumann S P,Nadlinger D P}. In quantum computation \cite{DiVincenzo D P,Nielsen M A}, recent protocols of Refs. \cite{Liu B J,Shi Z C,Kang Y H,Liu S} show that quantum computation can be implemented fast and robustly via different techniques, such as geometric phases, reverse engineering and composite pulses. As a quantum communication scheme, the dense coding resources and protocols have been investigated by many physics and information scientists \cite{Mattle K,Guo H,Meher N,Liu B H,Shaukat M I}.

For a closed system not to interact with its surroundings environment, quantum entanglement will not decay with time. However, for an open quantum system, decoherence \cite{Rijavec S} and dissipation always occur due to the unavoidable interaction with its surrounding environment, entanglement always undergoes either asymptotic degradation or sudden death processes \cite{Yu T}. Therefore, how to effectively protect and generate quantum entanglement has become a very important topic in open quantum systems.

\begin{figure}[tbp]
    \includegraphics[width=7cm,height=5cm]{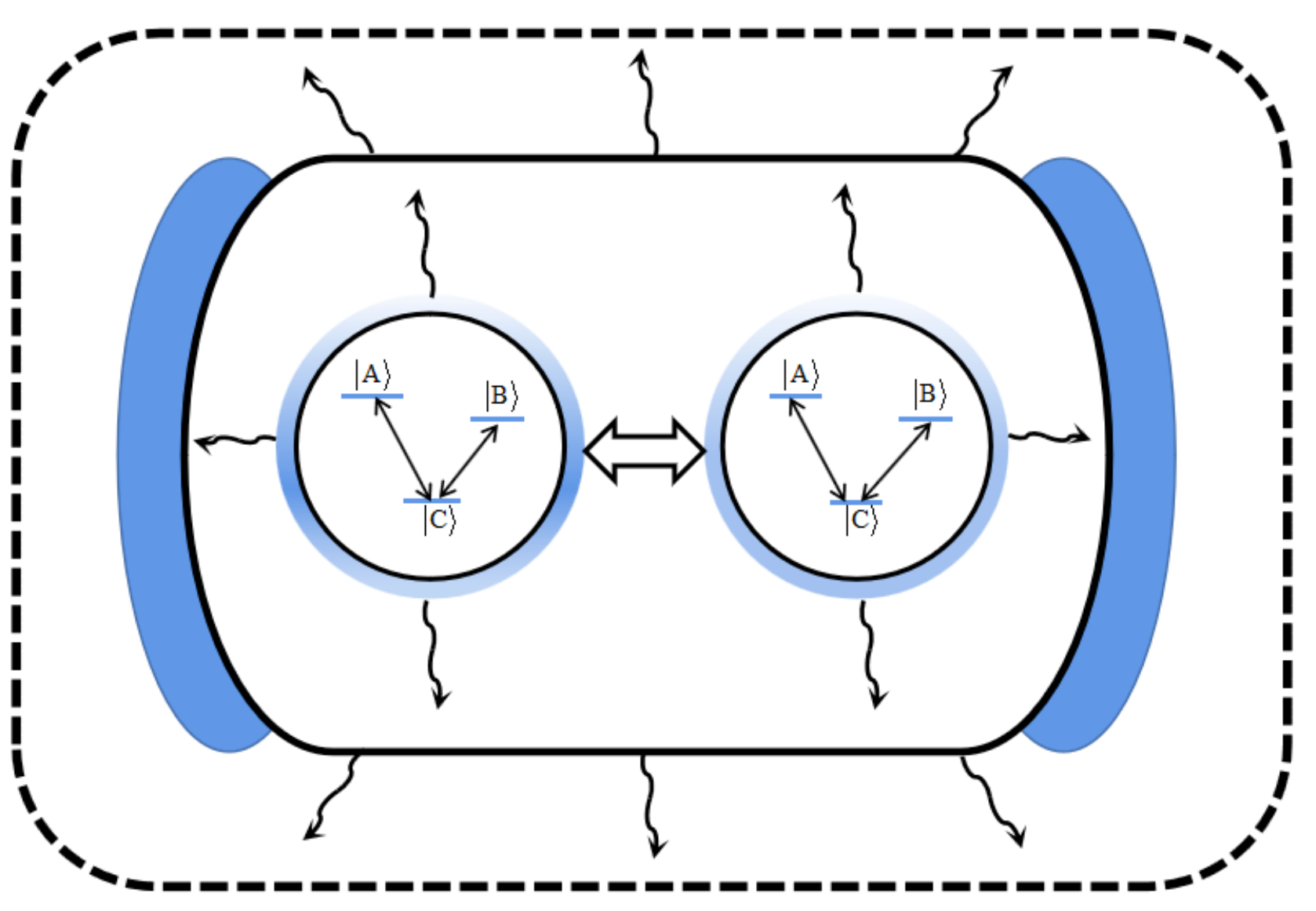}
	\caption{(Color online) Two identical V-type qutrit atomic system interacts with a dissipative cavity, which couples with an external environment. At the same time, there is the dipole-dipole interaction between the two atoms. The transition frequencies from atomic ground state $|C\rangle $ to excited state $|A\rangle $, $|B\rangle $ is $\omega _{A}$, $\omega _{B}$, respectively.}
	\label{fig:1}
\end{figure}

For nearly two decades, a great deal of attention has been devoted to the experimental and theoretical studies of entanglement dynamics of open qubit systems and some effective methods of protecting and regulating entanglement have also been proposed. For example, the memory and feedback effect of non-Markovian reservoirs can reduce the decay of entanglement and enhance the entanglement revival \cite{Zou,Zou H M,Mu Q}. The authors in Refs. \cite{Faizi E,Zhang Y J,Flores M M} found that the initial entanglement between the two qubits can be preserved by using the additional qubits. The results in Refs. \cite{Mortezapour A,Wang Q,Gholipour H} showed that the classical driving can effectively protect and regulate the quantum entanglement. The quantum Zeno effect can be used to produce and expand quantum entanglement \cite{Zaffino R L,Wang X B,Liu R}. The weak measurement can also protect entanglement from decoherence \cite{Kim Y S,Wang S C,Liao X P}. The feedback control can not only minimize the loss of entanglement between the two qubits but also create and stabilize the highly entangled states \cite{Zhang J,Rafiee M,Liu Z}. Besides these, the dipole-dipole interaction between atoms can also protect the entanglement from sudden death \cite{Altintas F,Fasihi}.

In recent years, open qutrit systems are also concerned and the methods of protecting entanglement in open qubit systems have also been used to the qutrit systems because the latter is superior physical resources than the former in quantum information processing. For instance, the auxiliary quantum qutrits \cite{Ahansaz B}, the classical driving field \cite{Metwally N}, the weak measurement and the measurement reversal \cite{Xiao X,Wang M J,Li W J,Xia Y J} are also effective ways in protecting and regulating entanglement of qutrit systems. In addition, a lot of progress has been made in the experimental implementation of  realistic systems \cite{Bronn N T,Vlastakis B,McKay D C,Hettrich M,Mok W K,Takahashi H,Fink J M,Leek P J,Stute A}. These works inspire us to investigate the effect of dipole-dipole interaction on the entanglement dynamics of open qutrit systems. Our purpose is to understand whether the dipole-dipole interaction can also protect the entanglement of qutrit systems. We find that the dipole-dipole interaction can not only protect and generate entanglement very effectively, but also enhance the regulation effect of spontaneously generated interference (SGI) parameters on entanglement for the initially partial entangled and product states, which will provide some references in the theoretical and experimental research of open qutrit systems.

In this paper, we investigated the entanglement dynamics of the two V-type atoms with dipole-dipole interaction in dissipative cavity. We can see that the SGI parameter has different effects on entanglement dynamics under different initial states, the entanglement dynamics has also obvious differences under the weak or strong cavity-environment coupling conditions, and the dipole-dipole interaction can regulate the entanglement dynamics behaviours.

The work is organized as follows. In Section II, we structure a physical model and obtain its analytical solution. In Section III, we give the entanglement negativity between two V-type atoms. Results and discussions are given in Section IV. And finally, the paper is ended with a brief conclusion in Section V.

\section{ Physical model}

\ We study a coupled system of two identical V-type atoms with dipole-dipole interaction in a dissipative single-mode cavity, which couples with an external environment considered as a set of continuum
harmonic oscillators, as depicted in Fig.1. Each atom has two excited states $|A\rangle $
and $|B\rangle $ which can spontaneously decay into the ground states $|C\rangle $ with the transition frequencies $\omega _{A} $ and $\omega _{B} $, respectively. Under the rotating-wave approximation \cite{Agarwal G S}, and in units of $\hbar=1$, the Hamiltonian of the total system is given by
\begin{equation} \label{EB201}
\hat{H}=\hat{H}_{A}+\hat{H}_{A F}+\hat{H}_{F E}+\hat{V}.
\end{equation}
The Hamiltonian $\hat{H}_{A}$ of two atoms is
\begin{equation}\label{EB202}
 \hat{H}_{A}=\sum_{l=1}^{2} \sum_{m=A, B} \omega _{m} \hat{\sigma}_{m}^{l+} \hat{\sigma}_{m}^{l-},
\end{equation}
where $\hat{\sigma}_{m}^{l\pm}$ ($m\mathcal{}$=$\mathit{A} $,$\mathit{B}$) is the raising and lowering operators of the $m\mathcal{}$-th excited state of the $l\mathcal{}$-th atom and $\hat{\sigma}_{m}^{l+}=\left|m\right\rangle _{l}\left\langle C\right|$, $\hat{\sigma}_{m}^{l-}=|C\rangle  _{l}^{}\langle m|$ \cite{Behzadi N,Li C}. At the same time, $\omega _{m} $  is the transition frequency of the $m\mathcal{}$-th excited state of atom. The Hamiltonian $\hat{H}_{A F}$ of atom-cavity interaction is
\begin{equation}\label{EB2022}
\hat{H}_{A F}=\sum_{l=1}^{2} \sum_{m=A, B}\left(g_{m} \hat{\sigma}_{m}^{l+} \hat{a}+g_{m}^{*} \hat{\sigma}_{m}^{l-} \hat{a}^{\dagger}\right),
\end{equation}
where $g_{m} $ is the coupling strength  between the $m\mathcal{}$-th excited state and the single-mode cavity, and $\hat{a} (\hat{a} ^{\dagger} )$ is the annihilation (creation) operators of single-mode cavity. The Hamiltonian $\hat{H}_{F E}$ of dissipative cavity is given by
\begin{equation}\label{EB2022}
\begin{aligned}
\hat{H}_{F E}=& \omega_{c} \hat{a}^{\dagger} \hat{a}+\int_{0}^{\infty} \eta \hat{B}^{\dagger}(\eta) \hat{B}(\eta) d \eta \\
&+\int_{0}^{\infty}\left\{G(\eta) \hat{a}^{\dagger} \hat{B}(\eta)+H_{\cdot }C_{\cdot }  \right\}d\eta ,
\end{aligned}
\end{equation}
where $\omega _{c}$ is the vibrational frequency of single-mode cavity, $\hat{B}^{\dagger}(\eta)$ and $\hat{B}(\eta)$ are creation and annihilation operators of environment in the mode $\eta$, and they follow the commutation relation $\left[\hat{B}(\eta), \hat{B}^{\dagger}\left(\eta^{\prime}\right)\right]=\delta\left(\eta-\eta^{\prime}\right)$.
Otherwise, $G(\eta )=\sqrt{\kappa /\pi }$ is the coupling coefficient of cavity-environment and $\kappa$ is the decay rate of the cavity, $H_{\cdot }C _{\cdot } $ is Hermitian conjugation. And the dipole-dipole interaction $\hat{V}$ between the two atoms can be written as
\begin{equation}\label{EB2022}
\hat{V} =\Omega \sum_{m=A, B} \sum_{n=A, B}\left(\hat{\sigma}_{m}^{1+} \hat{\sigma}_{n}^{2-}+\hat{\sigma}_{m}^{1-} \hat{\sigma}_{n}^{2+}\right),
\end{equation}
where $\Omega $ is the coupling constant of dipole-dipole interaction between atoms \cite{Li K,Mandilara A}, and $\Omega=\frac{1}{4 \pi\epsilon \mathbf{} _{0}}\left[\mathbf{d}\cdot\mathbf{ d} -\frac{3\left(\mathbf{d}\cdot\mathbf{r} _{12}\right)\left(\mathbf{d }\cdot\mathbf{r} _{12}\right)}{r_{12}^{2}}\right] r_{12}^{-3}$, where $\mathbf{r}_{12}=\mathbf{r}_{1}-\mathbf{r}_{2}$ is the relative position vector of any two atoms and $\mathbf{d}$ represents the electric dipole moment of the qutrit. This typical dipole-dipole interaction has been previously used in some literatures \cite{Peng J S,Mojaveri B,Li Y,Ahmadi Z,Fasihi M A,Tavassoly M K}.

In order to diagonalize Hamiltonian $\hat{H}_{F E}$, we introduce a set of new annihilation and creation operators, i.e. $\hat{A}(\omega)$ and $\hat{A}^{\dagger}(\omega)$ \cite{Nourmandipour A}. This new annihilation operator according to Fano theorem is written as
\begin{equation} \label{EB201}
\hat{A}(\omega)=\alpha (\omega) \hat{a}+\int \beta(\omega, \eta) \hat{B}(\eta) d \eta,
\end{equation}
where $\hat{A}(\omega)$ and its conjugate operator $\hat{A}^{\dagger}\left(\omega^{\prime}\right)$ satisfy the following commutation relation
\begin{equation} \label{EB201}
\left[\hat{A}(\omega), \hat{A}^{\dagger}\left(\omega^{\prime}\right)\right]=\delta\left(\omega-\omega^{\prime}\right),
\end{equation}
 where $\alpha(\omega)=\frac{\sqrt{\kappa  / \pi}}{\omega-\omega_{c}+i \kappa }$. Therefore the total Hamiltonian of the system can be written by using $\hat{A} (\omega )$ as
\begin{equation} \label{EB201}
\hat{H} =\hat{H}_{0}+\hat{H}_{I},
\end{equation}
where the free Hamiltonian $\hat{H}_{0}$ is
\begin{equation} \label{EB201}
\hat{H}_{0}=\sum_{l=1}^{2} \sum_{m=A, B} \omega_{m} \hat{\sigma }_{m}^{l+} \hat{\sigma }_{m}^{l-}+\int \omega \hat{A}^{\dagger}(\omega) \hat{A}(\omega) d \omega,
\end{equation}
and the interaction Hamiltonian $\hat{H}_{I}$ is
\begin{equation} \label{EB201}
\begin{aligned}
\hat{H}_{I}=& \sum_{l=1}^{2} \sum_{m=A, B} \int\left(g_{m} \alpha^{*}(\omega) \hat{\sigma}_{m}^{l+} \hat{A}(\omega)+H_{\cdot }  C_{\cdot } \right) d \omega \\
&+\Omega \sum_{m=A, B} \sum_{n=A, B}\left(\hat{\sigma}_{m}^{1+} \hat{\sigma}_{n}^{2-}+\hat{\sigma}_{m}^{1-}\hat{\sigma}_{n}^{2+}\right).
\end{aligned}
\end{equation}
The Schrodinger equation in the interaction picture is
\begin{equation} \label{EB201}
i \frac{d}{d t}|\psi(t)\rangle={\hat{H} _{int}   }(t)|\psi(t)\rangle,
\end{equation}
where $\hat{H}_{int}(t)= e^{i\hat{H}_{0}t}\hat{H}_{I}e^{-i\hat{H}_{0}t}$.

Supposing that the initial state of the total system is
\begin{equation} \label{EB201}
|\psi(0)\rangle=\left(C_{2}^{A}(0)\left|C_{1}, A_{2}\right\rangle+C_{1}^{B}(0)\left|B_{1}, C_{2}\right\rangle\right)_{S} \otimes|0\rangle_{E},
\end{equation}
where $\left|C_{2}^{A}(0)\right|^{2}+\left|C_{1}^{B}(0)\right|^{2}=1$, $|0\rangle_{E}$ denotes that the reservoir is in the vacuum state. Supposing the time evolution state $| \psi (t)\rangle$ is
\begin{equation} \label{EB201}
\begin{aligned}
|\psi(t)\rangle &=\left(C_{1}^{A}(t)\left|A_{1}, C_{2}\right\rangle+C_{1}^{B}(t)\left|B_{1}, C_{2}\right\rangle\right)_{S} \otimes|0\rangle_{E} \\
&+\left(C_{2}^{A}(t)\left|C_{1}, A_{2}\right\rangle+C_{2}^{B}(t)\left|C_{1}, B_{2}\right\rangle\right)_{S} \otimes|0\rangle_{E} \\
&+\int C_{\omega}(t)\left|C_{1}, C_{2}\right\rangle_{S}|1_{\omega } \rangle_{E} d \omega,
\end{aligned}
\end{equation}
here $\left|A_{1},C_{2}\right\rangle_{S}$ ($\left|B_{1},C_{2}\right\rangle_{S}$) represents that the first atom is in the excited state $\left|A\right\rangle$ ($\left|B\right\rangle$) and the other is in the ground states $\left|C\right\rangle$. Moreover, $\left|1_{\omega}\right\rangle_{E}$ indicates that the reservoir has only one excitation in the mode with frequency $\omega$. Let that the reservoir has the Lorentzian spectral density as
\begin{equation} \label{EB201}
J_{m n}(\omega)=\frac{1}{2 \pi} \frac{\gamma_{m n} \kappa ^{2}}{\left(\omega-\omega_{c}\right)^{2}+\kappa ^{2}},
\end{equation}
where $\gamma _{m n}=\frac{2 g_{m} g_{n}^{*}}{\kappa }$ is the relaxation rate of the excited state, and
\begin{equation} \label{EB201}
\gamma _{mm} =\gamma _{m} ,
\end{equation}
\begin{equation} \label{EB201}
\gamma_{m n}=\sqrt{\gamma_{m} \gamma _{n}} \theta, m \neq n,|\theta| \leq 1,
\end{equation}
where $\theta$ is defined as the SGI (spontaneously generated interference) parameter between the two decay channels $|A\rangle \to  |C\rangle$ and $|B\rangle \to  |C\rangle$ of each atom. The parameter $\theta$ depends on the angle between two dipole moments of the mentioned transitions. That is to say, $\theta = 0$ if the two dipole moments of the transitions are perpendicular to each other, which means that there is no SGI between the two decay channels. $\theta = 1$ if the two dipole moments are parallel, which is corresponding to the strongest SGI between the two decay channels.

Let that the two upper atomic states are degenerated and the atomic transitions are in resonant with the central frequency of the reservoir, i.e $\omega _{A} =\omega _{B} =\omega _{c}$, and $\gamma _{A}=\gamma _{B} =\gamma _{0} $ and $\gamma _{AB} =\gamma _{BA} =\gamma _{0} \theta$, $\gamma _{0}$ is the decay coefficient of the atomic excited state. Here, $\gamma_{0}/\kappa<<1/2$ is the week coupling regime and $\gamma_{0}/\kappa>>1/2$ is the strong coupling regime \cite{Bellomo B}.

Solving equation (11), we can obtain the probability amplitudes (see appendix)
\begin{equation} \label{EB201}
C_{l}^{A} (t)=\frac{1}{2} (C_{l}^{+}(t)+C_{l}^{-}(t)),
\end{equation}
\begin{equation} \label{EB201}
C_{l}^{B} (t)=\frac{1}{2} (C_{l}^{+}(t)-C_{l}^{-}(t)),
\end{equation}
here
\begin{equation} \label{EB201}
C_{1}^{\pm}(t)=\mathcal{G_{\pm}}(t) C_{1}^{\pm}(0)-\frac{e^{-2 i \Omega t}-\mathcal{G_{\pm}}(t)}{2}\left(C_{2}^{\pm}(0)-C_{1}^{\pm}(0)\right),
\end{equation}
\begin{equation} \label{EB201}
C_{2}^{\pm}(t)=\mathcal{G_{\pm}}(t) C_{2}^{\pm}(0)-\frac{e^{-2 i \Omega t}-\mathcal{G_{\pm}}(t)}{2}\left(C_{1}^{\pm}(0)-C_{2}^{\pm}(0)\right),
\end{equation}
where
\begin{equation} \label{EB201}
\begin{array}{l}
\mathcal{G_{\pm}}(t)=e^{-(\kappa +2 i \Omega) t / 2}
\\\times \left\{\cosh \left(\frac{D^{\pm} t}{2}\right)+\frac{\kappa -2 i \Omega}{D^{\pm}} \sinh \left(\frac{D^{\pm} t}{2}\right)\right\},
\end{array}
\end{equation}
and
\begin{equation} \label{EB201}
D^{\pm}=\sqrt{(\kappa +2 i \Omega)^{2}-4\left(2 i \Omega \kappa +\gamma _{0} \kappa (1 \pm \theta)\right)}.
\end{equation}

\section{entanglement negativity between two qutrits}
In this section, we consider the entanglement dynamics of two V-type qutrit systems in a common dissipative cavity. Firstly, we work out the explicit expressions of all probability amplitudes. From equations (17)-(20), we know that all probability amplitudes take the following simple forms
\begin{equation} \label{EB201}
\begin{aligned}
C_{1}^{A}(t)=\mathcal{Q}_{1}(t)\left(C_{1}^{A}(0)+C_{2}^{A}(0)\right)
\\+\mathcal{Q}_{2}(t)\left(C_{1}^{B}(0)+C_{2}^{B}(0)\right)
\\+\mathcal{Q}_{3}(t)\left(C_{1}^{A}(0)-C_{2}^{A}(0)\right)
\end{aligned},
\end{equation}
\begin{equation} \label{EB201}
\begin{aligned}
C_{2}^{A}(t)=\mathcal{Q}_{1}(t)\left(C_{2}^{A}(0)+C_{1}^{A}(0)\right)
\\+\mathcal{Q}_{2}(t)\left(C_{2}^{B}(0)+C_{1}^{B}(0)\right)
\\+\mathcal{Q}_{3}(t)\left(C_{2}^{A}(0)-C_{1}^{A}(0)\right)
\end{aligned},
\end{equation}
\begin{equation} \label{EB201}
\begin{aligned}
C_{1}^{B}(t)=\mathcal{Q}_{2}(t)\left(C_{1}^{A}(0)+C_{2}^{A}(0)\right)
\\+\mathcal{Q}_{1}(t)\left(C_{1}^{B}(0)+C_{2}^{B}(0)\right)
\\+\mathcal{Q}_{3}(t)\left(C_{1}^{B}(0)-C_{2}^{B}(0)\right)
\end{aligned},
\end{equation}
\begin{equation} \label{EB201}
\begin{aligned}
C_{2}^{B}(t)=\mathcal{Q}_{2}(t)\left(C_{2}^{A}(0)+C_{1}^{A}(0)\right)
\\+\mathcal{Q}_{1}(t)\left(C_{2}^{B}(0)+C_{1}^{B}(0)\right)
\\+\mathcal{Q}_{3}(t)\left(C_{2}^{B}(0)-C_{1}^{B}(0)\right)
\end{aligned},
\end{equation}
where
\begin{equation} \label{EB201}
\mathcal{Q}_{1}(t)=\frac{\mathcal{G_{+}}(t)+\mathcal{G_{-}}(t)}{4},
\end{equation}
\begin{equation} \label{EB201}
\mathcal{Q}_{2}(t)=\frac{\mathcal{G_{+}}(t)-\mathcal{G_{-}}(t)}{4},
\end{equation}
\begin{equation} \label{EB201}
\mathcal{Q}_{3}(t)=\frac{e^{-2 i \Omega  t}}{2}.
\end{equation}

In order to analyze the entanglement dynamics between the two V-type atoms, the reduced density operator is obtained from equation (13) by tracing the environmental freedom degree :
\begin{equation} \label{EB201}
\begin{aligned}
\rho (t) &=Tr_{E}(|\psi(t)\rangle\langle\psi(t)|)\\
&=\sum_{m=A, B} \sum_{n=A, B} C_{1}^{n}(t) C_{1}^{m *}(t)\left|n_{1}, C_{2}\right\rangle\left\langle m_{1}, C_{2}\right| \\
&+\sum_{m=A, B} \sum_{n=A, B} C_{1}^{n}(t) C_{2}^{m *}(t)\left|n_{1}, C_{2}\right\rangle\left\langle C_{1}, m_{2}\right| \\
&+\sum_{m=A, B} \sum_{n=A, B} C_{2}^{n}(t) C_{1}^{m *}(t)\left|C_{1}, n_{2}\right\rangle\left\langle m_{1}, C_{2}\right| \\
&+\sum_{m=A, B} \sum_{n=A, B} C_{2}^{n}(t) C_{2}^{m *}(t)\left|C_{1}, n_{2}\right\rangle\left\langle C_{1}, m_{2}\right| \\
&+\int\left|C_{\omega}(t)\right|^{2}\left|C_{1}, C_{2}\right\rangle\left\langle C_{1}, C_{2}\right|d\omega.
\end{aligned}
\end{equation}
In the basis
\begin{equation} \label{EB201}
\begin{array}{l}
\left\{\left|A_{1}, A_{2}\right\rangle,\left|A_{1}, B_{2}\right\rangle,\left|A_{1}, C_{2}\right\rangle,\left|B_{1}, A_{2}\right\rangle,\right. \\
\left.\left|B_{1}, B_{2}\right\rangle,\left|B_{1}, C_{2}\right\rangle,\left|C_{1}, A_{2}\right\rangle,\left|C_{1}, B_{2}\right\rangle,\left|C_{1}, C_{2}\right\rangle\right\},
\end{array}
\end{equation}
we can get the reduced density matrix of two atoms
\begin{equation} \label{EB201}
\rho(t)=\left(\begin{array}{ccccccccc}
0 & 0 & 0 & 0 & 0 & 0 & 0 & 0 & 0 \\
0 & 0 & 0 & 0 & 0 & 0 & 0 & 0 & 0 \\
0 & 0 & u_{33} & 0 & 0 & u_{36} & u_{37} & u_{38} & 0 \\
0 & 0 & 0 & 0 & 0 & 0 & 0 & 0 & 0 \\
0 & 0 & 0 & 0 & 0 & 0 & 0 & 0 & 0 \\
0 & 0 & u_{63} & 0 & 0 & u_{66} & u_{67} & u_{68} & 0 \\
0 & 0 & u_{73} & 0 & 0 & u_{76} & u_{77} & u_{78} & 0 \\
0 & 0 & u_{83} & 0 & 0 & u_{86} & u_{87} & u_{88} & 0 \\
0 & 0 & 0 & 0 & 0 & 0 & 0 & 0 & u_{99}
\end{array}\right),
\end{equation}
where $u_{33}=\left|C_{1}^{A}(t)\right|^{2}$, $u_{66}=\left|C_{1}^{B}(t)\right|^{2}$, $u_{77}=\left|C_{2}^{A}(t)\right|^{2}$, $u_{88}=\left|C_{2}^{B}(t)\right|^{2}$, $u_{99}=|C(t)|^{2}$, $u_{36}=C_{1}^{A}(t) C_{1}^{B *}(t)$, $u_{37}=C_{1}^{A}(t) C_{2}^{A *}(t)$, $u_{38}=C_{1}^{A}(t) C_{2}^{B *}(t)$, $u_{67}=C_{1}^{B}(t)C_{2}^{A *}(t)$, $u_{68}=C_{1}^{B}(t) C_{2}^{B *}(t)$, $u_{78}=C_{2}^{A}(t) C_{2}^{B *}(t)$, and $u_{lj}=u_{jl}^{*}(l\ne j)$, $|C(t)|^{2}=1-\left|C_{1}^{A}(t)\right|^{2}-\left|C_{1}^{B}(t)\right|^{2}-\left|C_{2}^{A}(t)\right|^{2}-\left|C_{2}^{B}(t)\right|^{2}$.

According to $\left\langle i_{A},j_{B}\left|\rho^{T_{1}}\right|k_{A},l_{B}\right\rangle\equiv\left\langle k_{A},j_{B}|\rho|i_{A},l_{B}\right\rangle$ \cite{Vidal G}, the partial transposition matrix is expressed as
\begin{equation} \label{EB201}
{\begin{array}{l}
\rho^{T_{1}}(t)= \\
\left(\begin{array}{ccccccccc}
0 & 0 & 0 & 0 & 0 & 0 & 0 & 0 & u_{73}  \\
0 & 0 & 0 & 0 & 0 & 0 & 0 & 0 & u_{83}  \\
0 & 0 & u_{33} & 0 & 0 & u_{63} & 0 & 0 & 0 \\
0 & 0 & 0 & 0 & 0 & 0 & 0 & 0 & u_{76}  \\
0 & 0 & 0 & 0 & 0 & 0 & 0 & 0 & u_{86}  \\
0 & 0 & u_{36} & 0 & 0 & u_{66} & 0 & 0 & 0 \\
0 & 0 & 0 & 0 & 0 & 0 & u_{77} & u_{78} & 0 \\
0 & 0 & 0 & 0 & 0 & 0 & u_{87} & u_{88} & 0 \\
u_{37} & u_{38} & 0 & u_{67} & u_{68} & 0 & 0 & 0 & u_{99}
\end{array}\right).
\end{array} }
\end{equation}

In the studies of entanglement dynamics of quantum systems, there are many methods to measure quantum entanglement, such as the partial entropy entanglement \cite{Wen Q}, the relation of entropy entanglement \cite{Wong G}, the concurrence \cite{Li D,Wootters W K} and the negativity \cite{Werner R F,Calabrese P}. For two qubits spin systems, the concurrence proposed by Wootters \cite{Wootters W K} is widely used to quantify the entanglement. However, for high-dimension quantum systems, the negativity is a more convenient way to quantify quantum entanglement. Here, we use the negativity to quantify the entanglement of the two V-type qutrit system, which is defined as
\begin{equation} \label{EB201}
\mathcal{N} (\rho)=-2 \sum_{i} \lambda_{i},
\end{equation}
where $\lambda_{i}$ is the eigenvalues of partial transposition matrix $\rho^{T_{1}}$ in equation (34), and $0\le \mathcal{N} (\rho)\le 1$ , where $\mathcal{N} (\rho)=0$ shows that the two atoms are in the product state, which is also called as the disentangled state. $\mathcal{N} (\rho)=1$ indicates that the two atoms are in the maximal entangled state. According to the definitions of partial transposition matrix and the negativity, we can numerically analyse the entanglement dynamics in detail.

\section{Results and Discussions}

In this section, we will analyse the effect of $\theta$, $\left|\psi(0)\right\rangle $ and $\Omega$ on the entanglement negativity of two V-type atoms under weak coupling ( $ \gamma_{0}/\kappa =0.1 $ ) and strong coupling  ( $\gamma_{0}/\kappa =10 $ ) regimes, respectively.

\subsection{ Without dipole-dipole interaction ($\Omega=0$)}

In this subsection, we investigate the effect of cavity-environment coupling and SGI parameter on the entanglement negativity when the two atoms are in different initial states without the dipole-dipole interaction.

Fig.2 indicates the effects of SGI parameter on entanglement dynamics under the weak and strong coupling conditions when the two V-type atoms are initially in the maximal entangled state. In Fig.2a (i.e. under the weak coupling condition), when $\theta = 0$, the entanglement negativity decays monotonously to a steady value 0.2 from 1.0, as shown the blue dot-dashed line. With the increasing of $\theta$, the decay rate of entanglement negativity also increases. However, the steady values of entanglement are equal under the different SGI parameters. Moreover, Fig.2b shows the dynamical behavior of entanglement negativity under the strong coupling condition. From Fig.2b, we can see that the entanglement negativity firstly decays monotonously, then revives to about 0.5, finally oscillates damply to the steady value 0.2 due to the feedback and memory effects of the environment. And a bigger SGI parameter corresponds to a greater revival amplitude and a smaller period of oscillation. Hence, for the initially  maximally entangled state, the entanglement negativity decays monotonously under the weak coupling condition while it will oscillate damply under the strong coupling condition. Also, the entanglement negativity has the same steady value under different coupling conditions and SGI parameters.

\begin{figure}[tbp]
	\includegraphics[width=7cm,height=4cm]{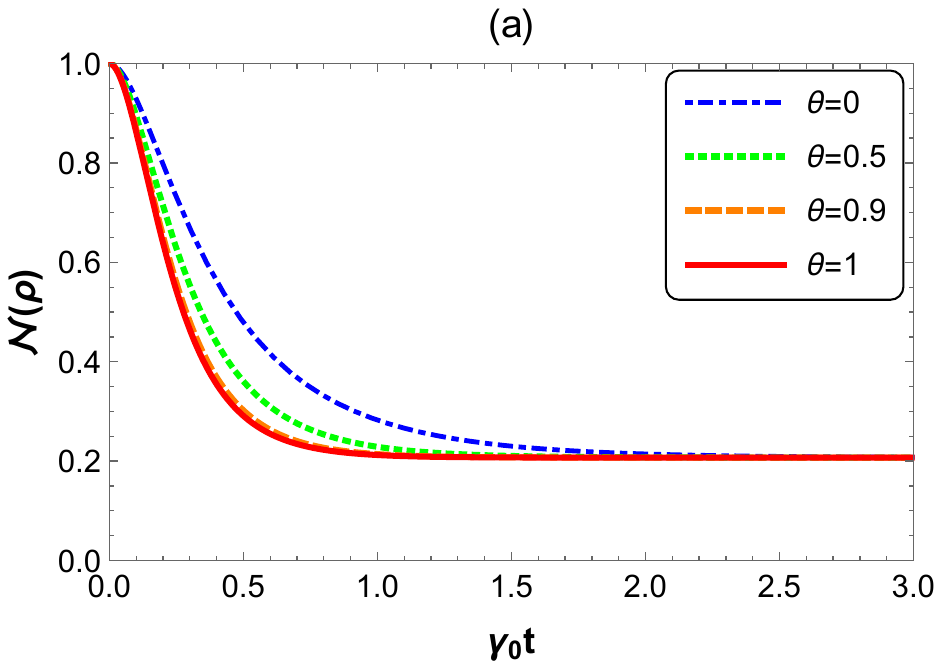}
	\includegraphics[width=7cm,height=4cm]{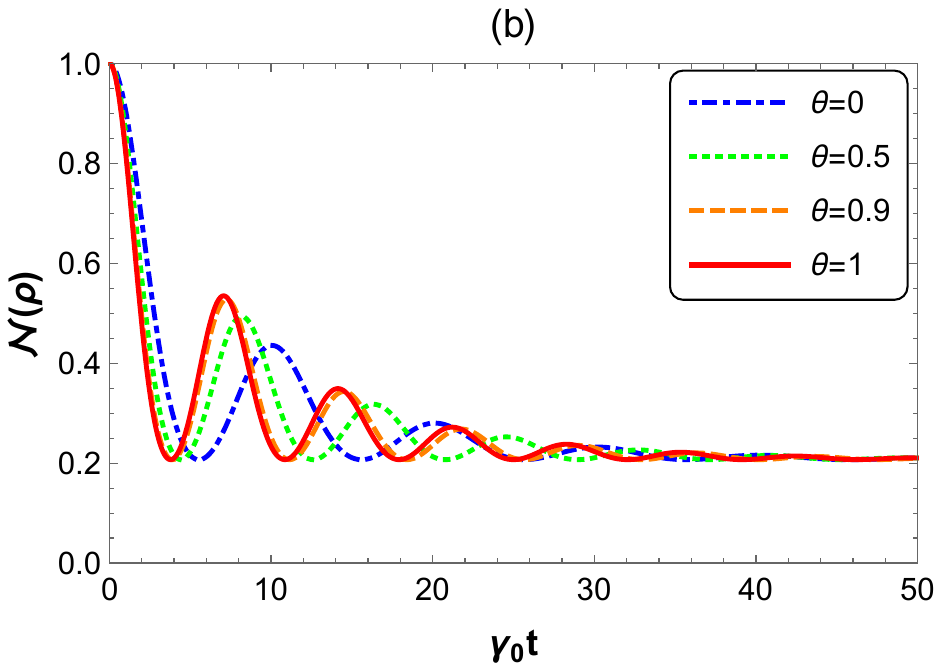}
	\caption{(Color online) The entanglement dynamics of two-qutrit system with $\theta=0$ (dot dashed line), $\theta=0.5$ (dot line), $\theta=0.9$ (dashed line) and $\theta=1$ (solid line). We assume that panel $\mathbf{a}$ is plotted under the weak coupling ($ \gamma_{0}/\kappa =0.1$ ), panel $\mathbf{b}$ under the strong coupling ($\gamma_{0}/\kappa =10$), and panels $\mathbf{a,b}$ with $\Omega=0$ Hz, $\gamma_{0}=1$. The initial state is determined by $\left|\psi (0)\right\rangle_{12}=\frac{\sqrt{2}}{2}(\left|C_{1},A_{2}\right\rangle+
\left|B_{1},C_{2}\right\rangle)_{S}\otimes\left|0\right\rangle_{E}$. }
	\label{fig:2}
\end{figure}

\begin{figure}[tbp]
	\includegraphics[width=7cm,height=4cm]{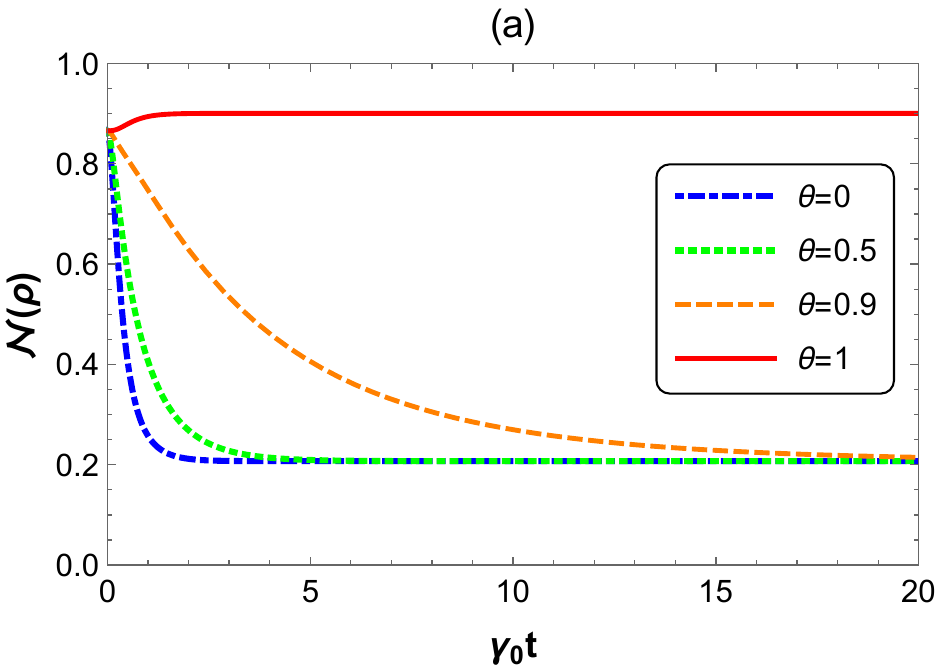}
	\includegraphics[width=7cm,height=4cm]{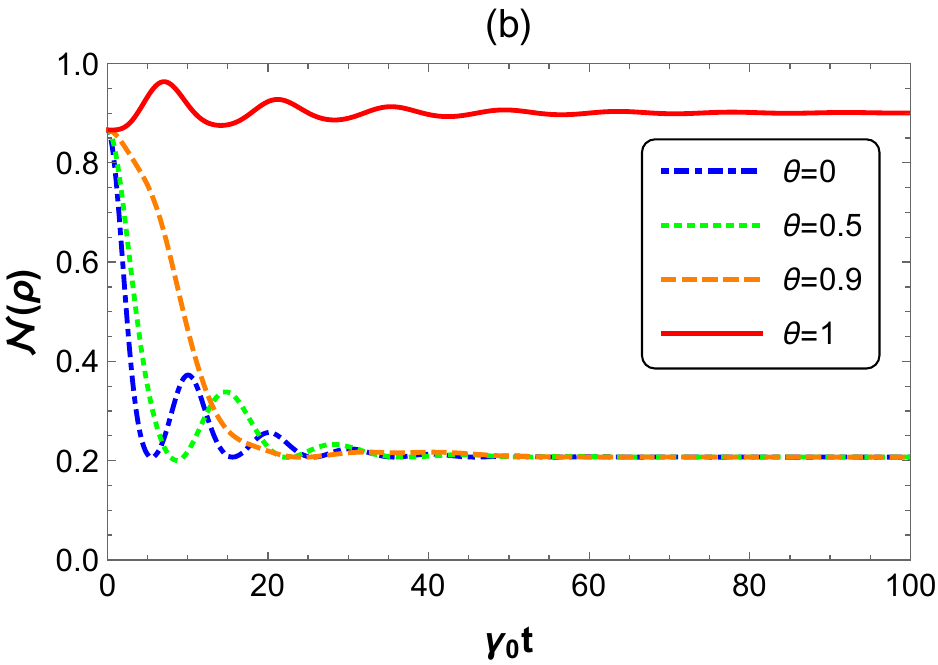}
	\caption{(Color online) The entanglement dynamics of two-qutrit system with $\theta=0$ (dot dashed line), $\theta=0.5$ (dot line), $\theta=0.9$ (dashed line) and $\theta=1$ (solid line). We assume that panel $\mathbf{a}$ is plotted under the weak coupling ($ \gamma_{0}/\kappa =0.1$ ), panel $\mathbf{b}$ under the strong coupling ($\gamma_{0}/\kappa =10$), and panels $\mathbf{a,b}$ with $\Omega = 0$ Hz, $\gamma_{0}=1$. The initial state is determined by $\left|\psi (0)\right\rangle_{12}=(-\frac{\sqrt{3}}{2} \left|C_{1},A_{2}\right\rangle+\frac{1}{2} \left|B_{1},C_{2}\right\rangle)_{S}\otimes\left|0\right\rangle_{E}$.}
	\label{fig:3}
\end{figure}

\begin{figure}[tbp]
	\includegraphics[width=7cm,height=4cm]{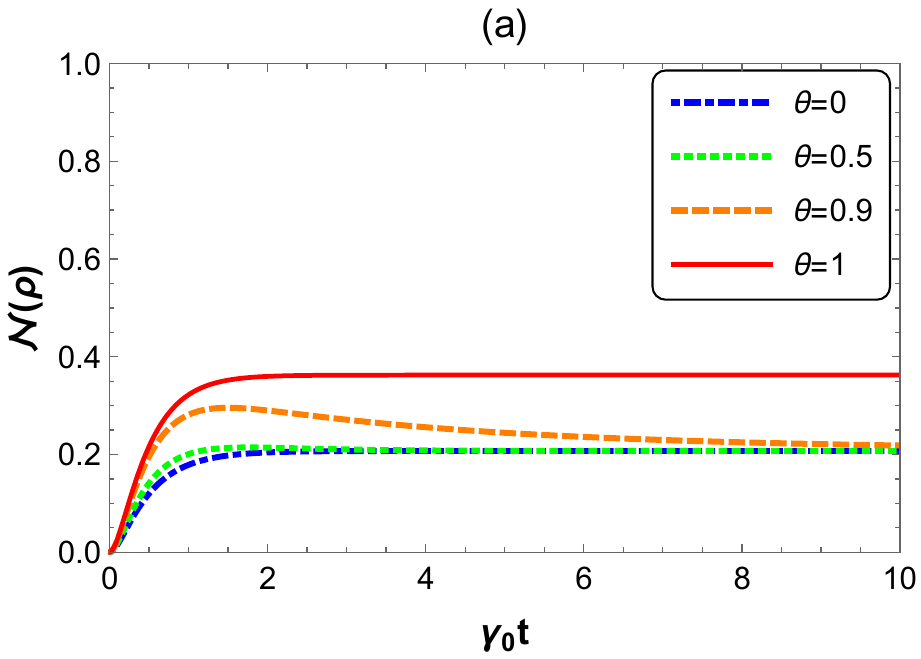}
	\includegraphics[width=7cm,height=4cm]{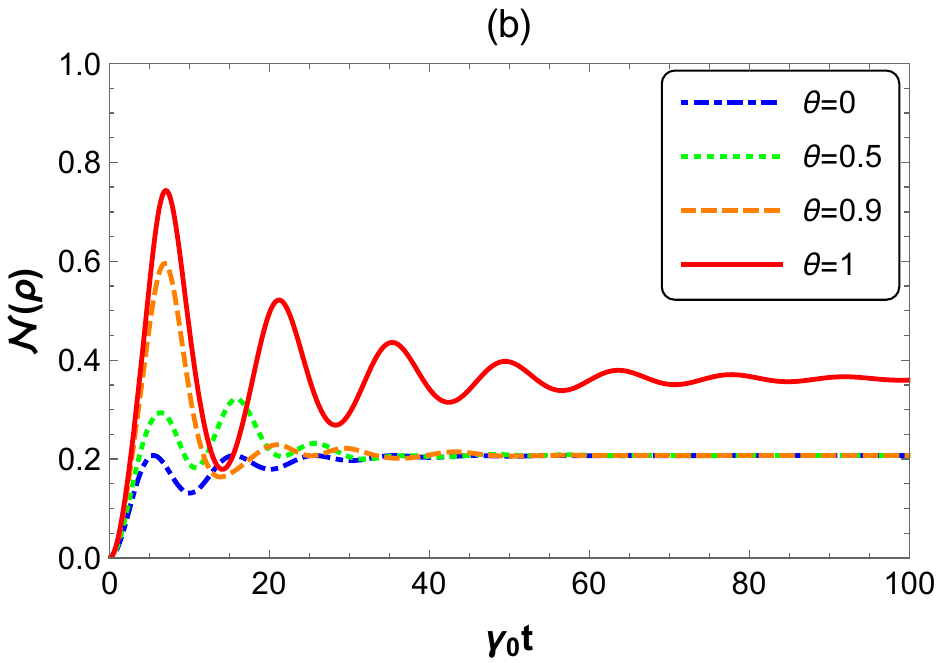}
	\caption{(Color online) The entanglement dynamics of two-qutrit system with $\theta=0$ (dot dashed line), $\theta=0.5$ (dot line), $\theta=0.9$ (dashed line) and $\theta=1$ (solid line). We assume that panel $\mathbf{a}$ is plotted under the weak coupling ($ \gamma_{0}/\kappa =0.1$ ), panel $\mathbf{b}$ under the strong coupling ($\gamma_{0}/\kappa =10$), and panels $\mathbf{a,b}$ with $\Omega = 0$ Hz, $\gamma_{0}=1$. The initial state is determined by $\left|\psi(0)\right\rangle_{12}=
\left|B_{1},C_{2}\right\rangle_{S}\otimes\left|0\right\rangle_{E}$.}
	\label{fig:4}
\end{figure}

Fig.3 shows the effect of SGI parameter on entanglement dynamics under different coupling conditions when the atoms are initially in the partially entangled state with the negativity 0.86. In Fig.3a (i.e. under the weak coupling condition), when $\theta <1$, the entanglement negativity also monotonously decays to a steady value 0.2 from 0.86 and the decay rate of entanglement negativity obviously become small with the increasing of $\theta$. In particular, when $\theta=1$, the entanglement negativity will grow to a new steady value 0.9 from 0.86, which is completely different from the case when $\theta<1$. Fig.3b shows the entanglement dynamics under strong coupling condition. From Fig.3b, it can be clearly seen that the revival-oscillation phenomena of entanglement negativity will occur under the strong coupling condition due to the interaction between the system and the environment. When $\theta<1$, the revival amplitude and the oscillation frequency of entanglement will become small with the SGI parameter increasing, but the steady values of entanglement are equal. As a result, when the SGI parameter $\theta<1$, the entanglement negativity also has the same steady value 0.2 in different coupling conditions, and a bigger SGI parameter corresponds to a smaller decay rate and revival amplitude of entanglement. However, when the SGI parameter $\theta=1$, the entanglement negativity will increase rather than decrease. Namely, the maximal SGI parameter is very beneficial for generating and protecting the quantum entanglement of the initially partially entangled state.

Fig.4 describes the effects of SGI parameter on entanglement dynamics under different coupling conditions when the atoms are initially in the disentangled state. From Fig.4, we can see that entanglement negativity increases to a steady value from zero, which is completely different from the cases of the maximal or partial entangled states (Fig.2 and Fig.3). The entanglement negativity increases monotonously under the weak coupling condition. However, under the strong coupling condition, the entanglement negativity firstly raises then oscillates damply to a steady value, and a bigger SGI parameter corresponds to a larger entanglement peak. In addition, the steady value of entanglement is equal to 0.2 when $\theta<1$ while it is equal to 0.36 when $\theta=1$. This denotes that the two atoms can be entangled though they are initially in the product state, which provides a method of preparing entangled states.

\subsection{ With dipole-dipole interaction ($\Omega\neq0$)}

In this subsection, we focus on the effect of dipole-dipole interaction between the two atoms on the entanglement negativity under different SGI parameters and cavity-environment coupling.

In Fig.5, we plot the curves of entanglement dynamics of two atoms in the initially maximal entangled state when the dipole-dipole interaction is equal to $3\gamma _{0}$ Hz (i.e. $\Omega=3\gamma _{0}$ Hz). Fig.5 shows that, if there is the dipole-dipole interaction between two atoms, the decay rate of entanglement negativity will decrease. In particular, under the strong coupling condition, the entanglement negativity can be protected very effectively, and it will tend to the steady value 0.2 only when the time is long enough (see Fig.5b). Moreover, the entanglement negativity decays faster when the SGI parameter increases. Therefore, the dipole-dipole interaction can protect very effectively the quantum entanglement for the initially maximal entangled state.

For the initially partial entangled state with $\mathcal{N} (\rho)=0.86$, the curves of entanglement dynamics of two atoms are given in Fig.6 when $\Omega=3\gamma _{0}$ Hz. From Fig.6a, we know that, under the weak coupling condition, the decay rate of entanglement negativity with $\Omega=3\gamma _{0}$ Hz is smaller than that with $\Omega=0$ Hz (see Fig.3a) when $\theta<1$, but the effect of dipole-dipole interaction on entanglement is small by comparing the red solid lines in Fig.3a and Fig.6a when $\theta=1$. Fig.6b tells us that, under the strong coupling condition, the entanglement negativity first increases from 0.86 for all values of $\theta$, and it will oscillate and tend to the steady value only when the time is long enough. The entanglement peak will become larger with the increase of $\theta$. Moreover, their steady values are 0.9 (when $\theta=1$) and 0.2 (when $\theta<1$), respectively. Hence, the dipole-dipole interaction can not only protect the entanglement very effectively, but also enhance the regulation effect of $\theta$ on entanglement for the initially partial entangled state.

\begin{figure}[tbp]
	\includegraphics[width=7cm,height=4.5cm]{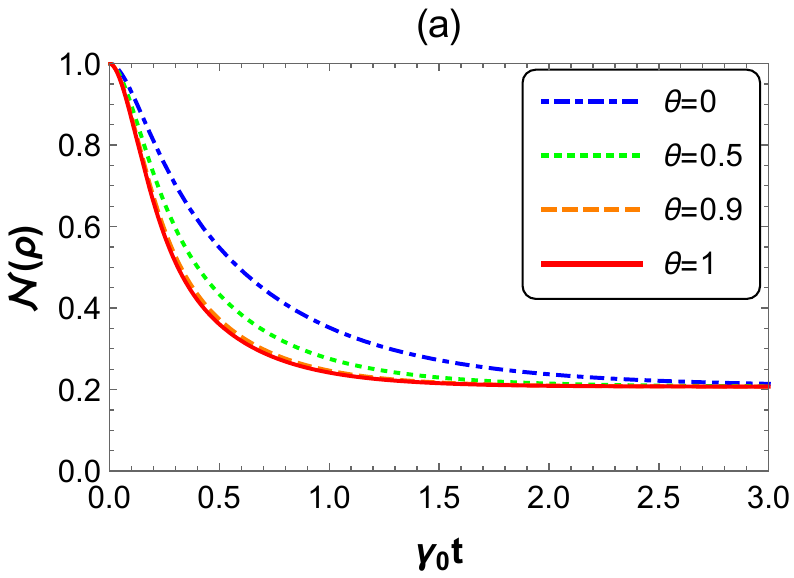}
	\includegraphics[width=7.5cm,height=5cm]{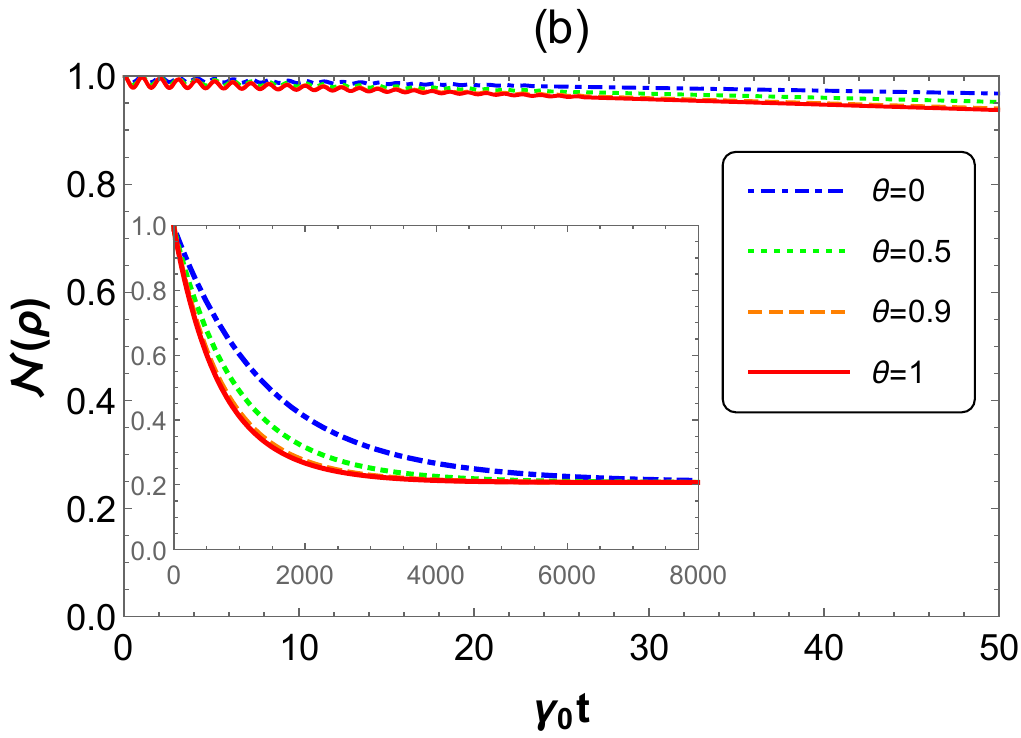}
	\caption{(Color online) The entanglement dynamics of two-qutrit system with $\theta=0$ (dot dashed line), $\theta=0.5$ (dot line), $\theta=0.9$ (dashed line) and $\theta=1$ (solid line). We assume that panel $\mathbf{a}$ is plotted under the weak coupling ($ \gamma_{0}/\kappa =0.1$ ), panel $\mathbf{b}$ under the strong coupling ($\gamma_{0}/\kappa =10$), and panels $\mathbf{a,b}$ with $\Omega = 3\gamma _{0}$ Hz, $\gamma_{0}=1$. The initial state is determined by $\left|\psi (0)\right\rangle_{12}=\frac{\sqrt{2}}{2}(\left|C_{1},A_{2}\right\rangle+
\left|B_{1},C_{2}\right\rangle)_{S}\otimes\left|0\right\rangle_{E}$. }
	\label{fig:5}
\end{figure}

\begin{figure}[tbp]
	\includegraphics[width=8cm,height=4.7cm]{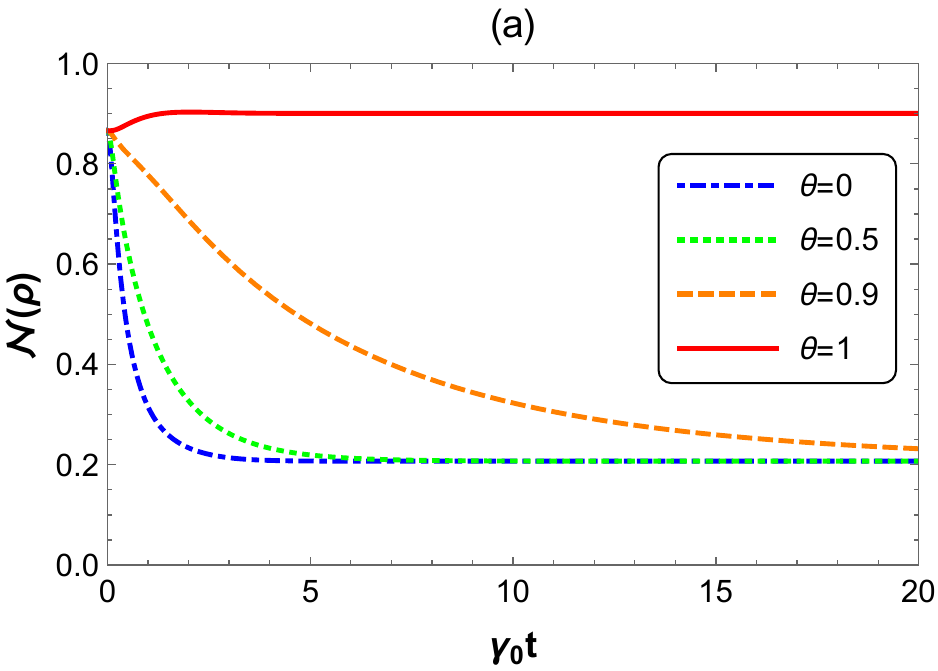}
	\includegraphics[width=8.5cm,height=4.8cm]{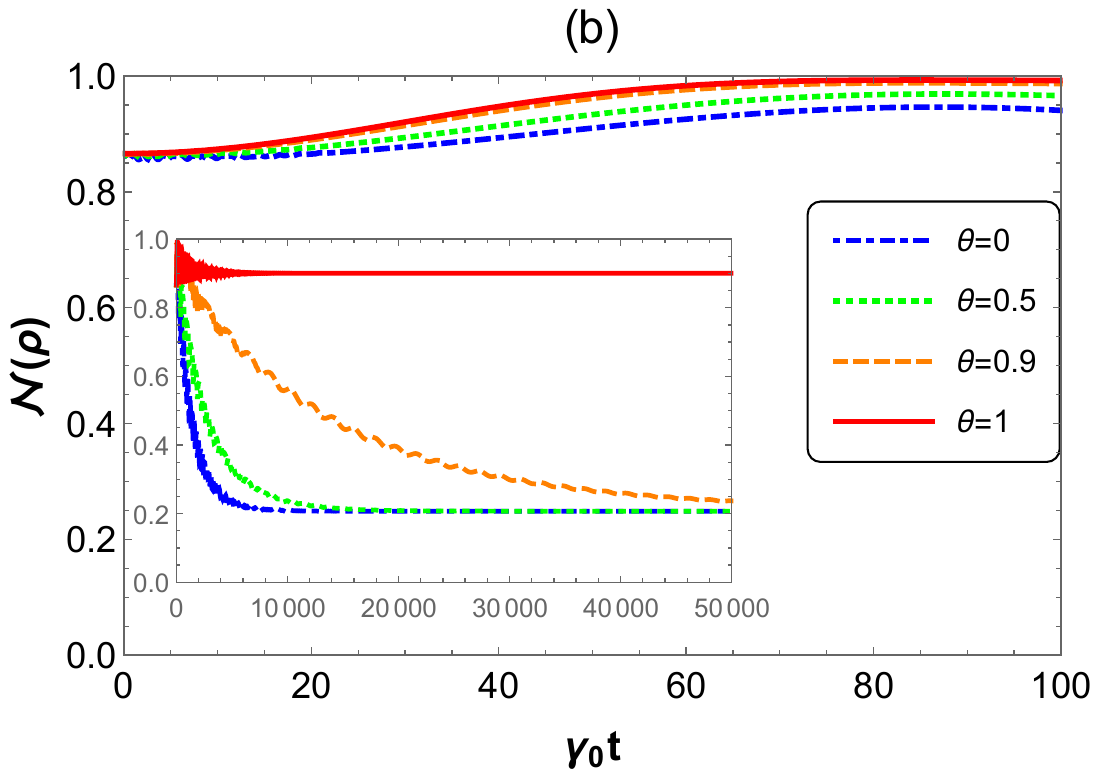}
	\caption{(Color online) The entanglement dynamics of two-qutrit system with $\theta=0$ (dot dashed line), $\theta=0.5$ (dot line), $\theta=0.9$ (dashed line) and $\theta=1$ (solid line). We assume that panel $\mathbf{a}$ is plotted under the weak coupling ($ \gamma_{0}/\kappa =0.1$ ), panel $\mathbf{b}$ under the strong coupling ($\gamma_{0}/\kappa =10$), and panels $\mathbf{a,b}$ with $\Omega = 3\gamma _{0}$ Hz, $\gamma_{0}=1$. The initial state is determined by $\left|\psi (0)\right\rangle_{12}=(-\frac{\sqrt{3}}{2} \left|C_{1},A_{2}\right\rangle+\frac{1}{2} \left|B_{1},C_{2}\right\rangle)_{S}\otimes\left|0\right\rangle_{E}$.}
	\label{fig:6}
\end{figure}

\begin{figure}[tbp]
	\includegraphics[width=7cm,height=4.5cm]{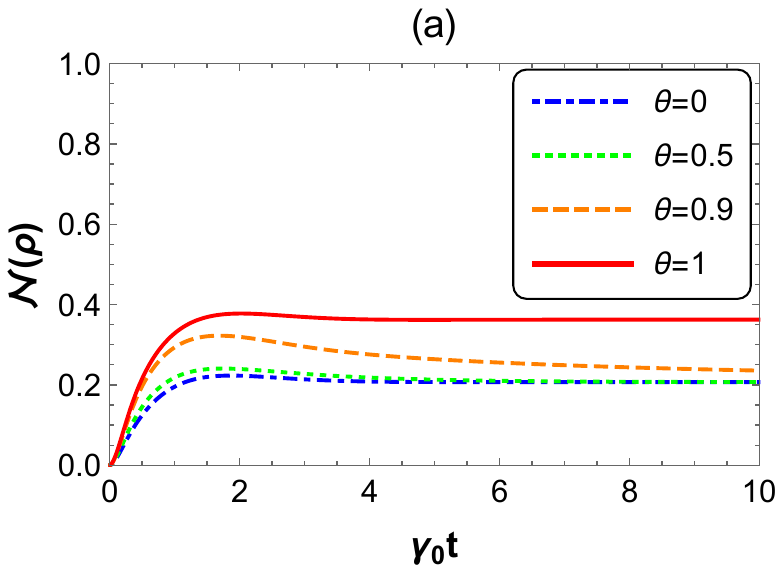}
	\includegraphics[width=7cm,height=4cm]{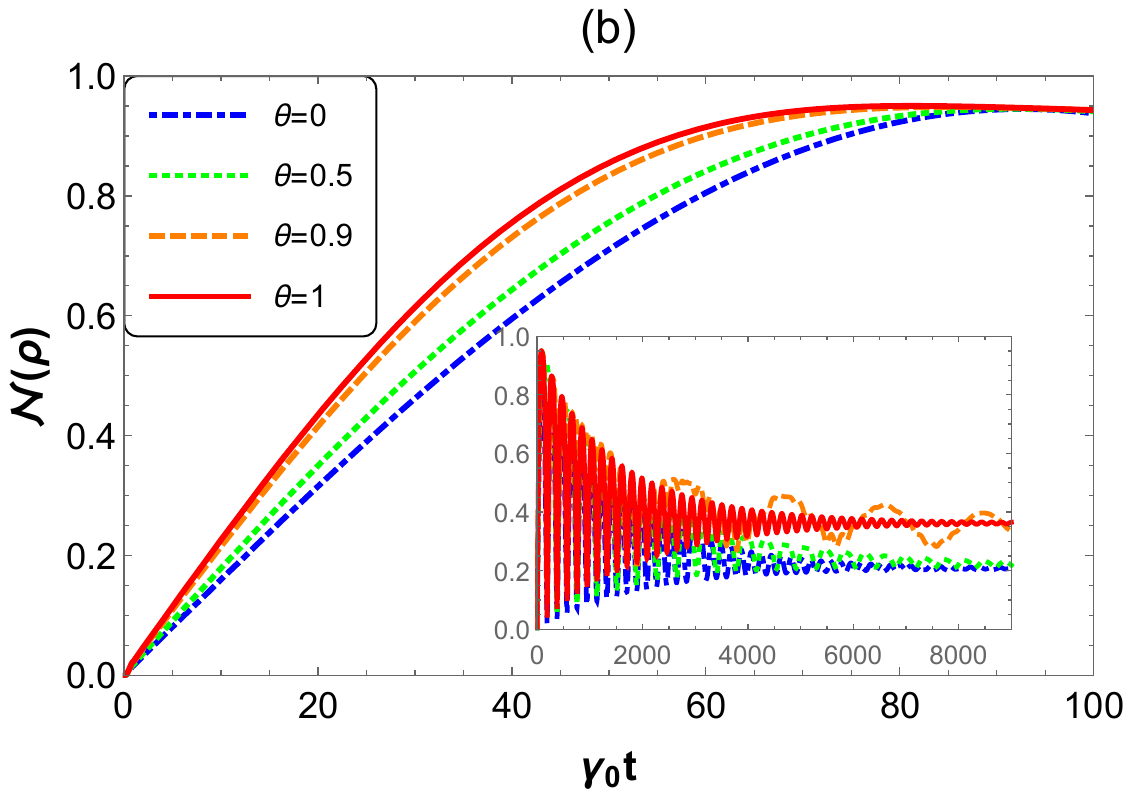}
	\caption{(Color online) The entanglement dynamics of two-qutrit system with $\theta=0$ (dot dashed line), $\theta=0.5$ (dot line),$\theta=0.9$ (dashed line) and $\theta=1$ (solid line). The initial state is determined by $\left|\psi(0)\right\rangle_{12}=
\left|B_{1},C_{2}\right\rangle_{S}\otimes\left|0\right\rangle_{E}$. We assume that panel $\mathbf{a}$ is plotted under the weak coupling ($ \gamma_{0}/\kappa =0.1$ ), panel $\mathbf{b}$ under the strong coupling ($\gamma_{0}/\kappa =10$), and panels $\mathbf{a}$ , $\mathbf{b}$ with $\Omega=3\gamma _{0}$ Hz, $\gamma_{0}=1$.}
	\label{fig:7}
\end{figure}

\begin{figure}[tbp]
	\includegraphics[width=7cm,height=3.5cm]{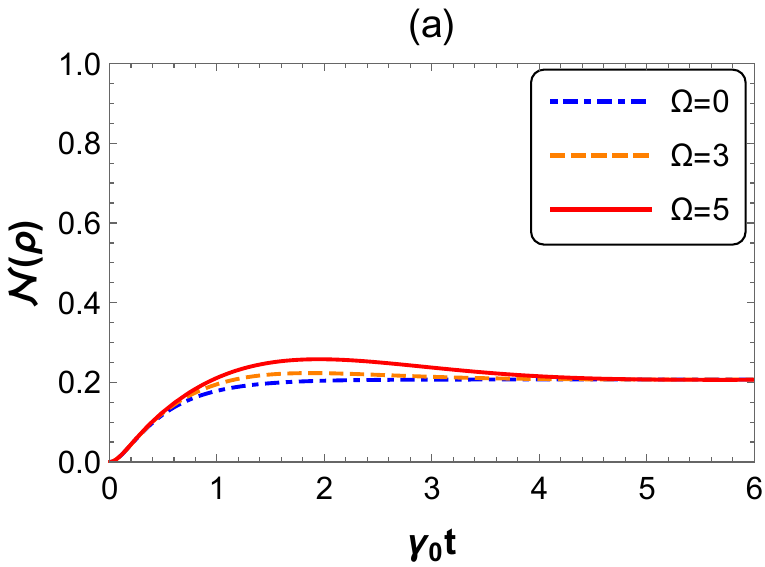}
	\includegraphics[width=7cm,height=3.5cm]{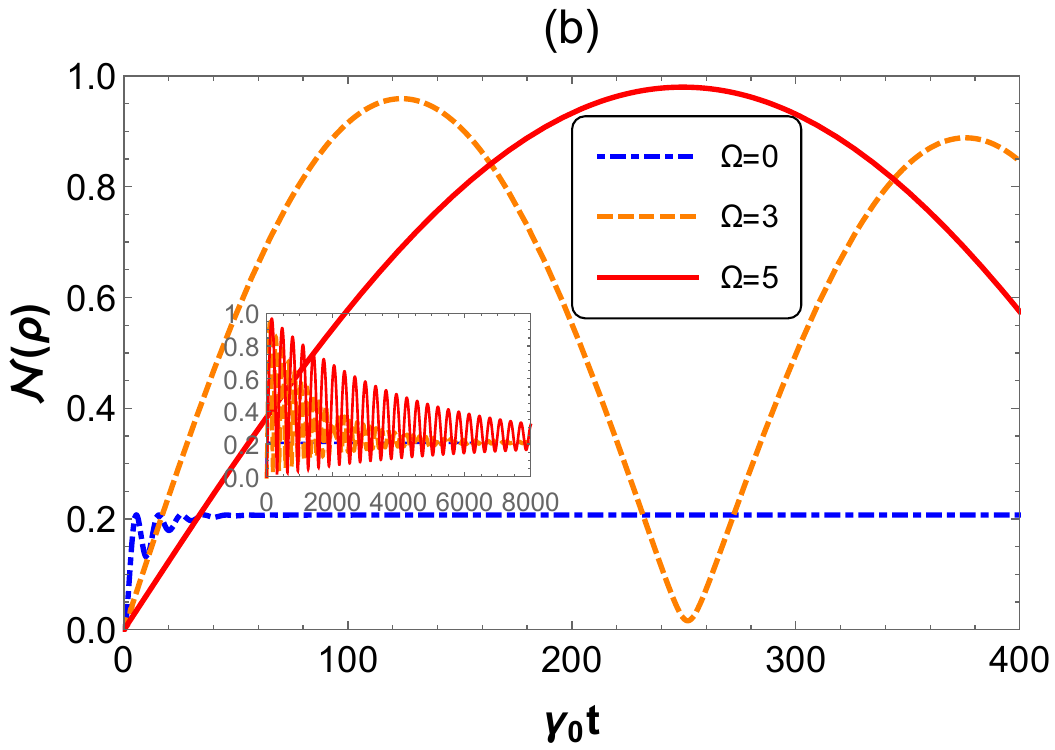}
	\caption{(Color online) The entanglement dynamics of two-qutrit system with $\Omega=0$ Hz (dot dashed line), $\Omega=3\gamma _{0}$ Hz (dashed line) and $\Omega=5\gamma _{0}$ Hz (solid line). The initial state is determined by $\left|\psi(0)\right\rangle_{12}=
\left|B_{1},C_{2}\right\rangle_{S}\otimes\left|0\right\rangle_{E}$. We assume that panel $\mathbf{a}$ is plotted under the weak coupling ($ \gamma_{0}/\kappa =0.1$ ), panel $\mathbf{b}$ under the strong coupling ($\gamma_{0}/\kappa =10$), and panels $\mathbf{a,b}$ with $\theta=0$, $\gamma_{0}=1$.}
	\label{fig:8}
\end{figure}

For the initially product state, the entanglement dynamics of two atoms under the dipole-dipole interaction is displayed in Fig.7. Fig.7a indicates that the entanglement negativity will increase to the different steady values from zero under the weak coupling condition when $\Omega=3\gamma _{0}$ Hz, and a bigger SGI parameter corresponds to a greater maximal value. Under the strong coupling condition, the entanglement negativity firstly increases from zero to about 0.95 then oscillates damply to the steady value when $\Omega=3\gamma _{0}$ Hz and a bigger SGI parameter corresponds to a larger amplitude of oscillation, as shown Fig.7b. Furthermore, the steady values of entanglement negativity are respectively 0.36 (when $\theta=1$) and 0.2 (when $\theta<1$) for both weak and strong coupling conditions. Therefore, the two atoms not only can be entangled, but also can achieve a very large value of entanglement negativity when $\Omega=3\gamma _{0}$ Hz though they are initially in a product state.

To further demonstrate the effect of dipole-dipole interaction on entanglement dynamics, we draw the curves of entanglement dynamics of two atoms in the initially product  state under the different dipole-dipole interaction when $\theta=0$ (see Fig.8). From Fig.8a, we see that, under the weak coupling condition, the entanglement raises to 0.2 when $\Omega=0$ Hz while it will rise to a certain maximum and then decrease to 0.2 when $\Omega>0$ Hz. The greater the dipole-dipole interaction is, the greater the entanglement peak is. Fig.8b shows that, under the strong coupling condition, the dipole-dipole interaction plays a significant role in the regulation of entanglement. For example, the entanglement negativity firstly rises from zero and then oscillates to 0.2 when $\Omega=0$ Hz. However, the entanglement negativity will increase to about 1.0 from zero when $\Omega=5\gamma _{0}$ Hz. The amplitude and period of oscillation of entanglement negativity increase significantly with the increase of $\Omega$. Thus, the dipole-dipole interaction can enlarge the entanglement and prolong the entangled time.

We may give physical interpretations of the above results. Firstly, we discuss the effect of initial state on the entanglement dynamics. For the maximal entangled state, the entanglement negativity will reduce under the influence of cavity and environment. When the total system is initially in $\left|B_{1},C_{2}\right\rangle_{S}\left|0\right\rangle_{E}$, the entanglement negativity will increase from zero because the exciton of the first atom is exchanged to the second atom through the spontaneously generated interference. Secondly, we analyze the effect of the cavity-environment coupling on the entanglement dynamics. Under the weak coupling condition, the quantum information will continually dissipate to the environment, thus the entanglement negativity will reduce monotonously, as shown in Fig.2a, Fig.3a and Fig.4a. Under the strong coupling condition, the entanglement negativity will oscillate damply due to the feedback and the memory effects of environment, as shown in Fig.2b, Fig.3b and Fig.4b. Thirdly, we consider the effect of the SGI parameter on entanglement dynamics. If the two atoms are initially in the maximal entangled state, the decay rate of entanglement negativity will increase with the SGI parameter increasing because the probabilities that the atoms exchange quantum information with the cavity are approximately equal, as shown in Fig.2 and Fig.5. For the partial entangled state or the product state, the smaller the SGI parameter is, the smaller the angle between the two dipole moments of the atom is (i.e. the closer the two dipole moments are to parallel), and the easier the exciton is exchanged between the two atoms through the cavity mode. So the entanglement decays more slowly and even increases when $\theta=1$, as shown in Fig.3-4 and Fig.6-7. Finally, we discuss the effect of the dipole-dipole interaction on the entanglement dynamics. When there is the dipole-dipole interaction between the two atoms, the entanglement can be protected very effectively because quantum information can be better trapped between the two atoms. The greater the dipole-dipole interaction is, the faster and the more the quantum information is exchanged between the two atoms, thus the better the entanglement is protected and the more the entanglement is generated, as shown in Fig.5-8.

\section{Conclusion}

In this work, we study a coupled system of two V-type atoms with dipole-dipole interaction in a dissipative single-mode cavity, which couples with an external environment. Firstly, we diagonalize Hamiltonian of dissipative cavity by introducing a set of new creation and annihilation operators according to Fano theorem. Then, we obtain the analytical solution of this model by solving the time dependent Schrodinger equation. We also detailedly discuss the influences of cavity-environment coupling, SGI parameter and dipole-dipole interaction between the two atoms on entanglement dynamics under different initial states. We found that the entanglement will change monotonously under the weak coupling condition while it will oscillate under the strong coupling condition. For the initially maximal entangled state, the entanglement has the same steady value under different coupling conditions and SGI parameters. The bigger SGI parameter corresponds to the larger decay rate and revival amplitude of entanglement. For the initially partial entangled state, when the SGI parameter $\theta<1$, the entanglement will reduce to the same steady value under different coupling conditions, and a bigger SGI parameter corresponds to a smaller decay rate and revival amplitude of entanglement. However, when the SGI parameter $\theta=1$, the entanglement will increase rather than decrease. Namely, a bigger SGI parameter can very effectively protect and generate the entanglement for the initially partial entangled state. For the initially product state, if $\Omega=0$, the entanglement will rise to 0.2 from zero when $\theta<1$ and it increases to 0.36 from zero when $\theta=1$. When there is the dipole-dipole interaction between the two atoms, the entanglement can be protected very effectively, and the more entanglement will be generated. Also, the larger the dipole-dipole interaction is, the slower the entanglement decays and the more the entanglement is generated. So the dipole-dipole interaction can not only very effectively protect and generate entanglement, but also enhance the regulation effect of SGI parameter on entanglement. We also give the physical interpretations of all results.

\begin{acknowledgments}
This work was supported by the Doctoral Science Foundation of Hunan Normal University, China.
\end{acknowledgments}

\appendix
\section{equations (17)-(18)}
In this appendix, we intend to obtain the probability amplitudes. To begin with, we need to solve Schrodinger equation in the interaction picture which is
\begin{equation} \label{EB201}
i \frac{d}{d t}|\psi(t)\rangle={\hat{H} _{int}   }(t)|\psi(t)\rangle,
\end{equation}
where
\begin{equation} \label{EB201}
\begin{aligned}
&\hat{H}_{int}(t)\\
=& e^{i \hat{H}_{0} t} \hat{H}_{I} e^{-i \hat{H}_{0} t} \\
=& \sum_{l=1}^{2} \sum_{m=A, B} \int\left(g_{m} \alpha^{*}(\omega) \hat{\sigma}_{m}^{l+} \hat{A}(\omega) e^{i\left(\omega_{m}-\omega\right) t}+H.C.\right) d \omega \\
&+\Omega \sum_{m=A, B} \sum_{n=A,B}\left(\hat{\sigma}_{m}^{1+} \hat{\sigma}_{n}^{2-} e^{i\left(\omega_{m}-\omega_{n}\right) t}+H.C.\right).
\end{aligned}
\end{equation}
Supposing that the initial state of the total system is
\begin{equation} \label{EB201}
|\psi(0)\rangle=\left(C_{2}^{A}(0)\left|C_{1}, A_{2}\right\rangle+C_{1}^{B}(0)\left|B_{1}, C_{2}\right\rangle\right)_{S} \otimes|0\rangle_{E},
\end{equation}
where $\left|C_{2}^{A}(0)\right|^{2}+\left|C_{1}^{B}(0)\right|^{2}=1$, $|0\rangle_{E}$ denotes that the reservoir is in the vacuum state. Supposing the time evolution state $|\psi(t)\rangle$ is
\begin{equation} \label{EB201}
\begin{aligned}
|\psi(t)\rangle &=\left(C_{1}^{A}(t)\left|A_{1}, C_{2}\right\rangle+C_{1}^{B}(t)\left|B_{1}, C_{2}\right\rangle\right)_{S} \otimes|0\rangle_{E} \\
&+\left(C_{2}^{A}(t)\left|C_{1}, A_{2}\right\rangle+C_{2}^{B}(t)\left|C_{1}, B_{2}\right\rangle\right)_{S} \otimes|0\rangle_{E} \\
&+\int C_{\omega}(t)\left|C_{1}, C_{2}\right\rangle_{S}|1_{\omega } \rangle_{E} d \omega,
\end{aligned}
\end{equation}
here $\left|A_{1},C_{2}\right\rangle_{S}$ ($\left|B_{1},C_{2}\right\rangle_{S}$) represents that the first atom is in the excited state $\left|A\right\rangle$ ($\left|B\right\rangle$) and the other is in the ground states $\left|C\right\rangle$. Moreover, $\left|1_{\omega}\right\rangle_{E}$ indicates that the reservoir has only one excitation in the mode with frequency $\omega$. Substituting equations (A2) and (A4) into equation (A1), the differential equations can been obtained as
\begin{equation} \label{EB201}
\begin{array}{l}
\dot{C}_{l}^{m}(t)=-i g_{m} \int e^{i\left(\omega_{m}-\omega\right) t} \alpha^{*}(\omega) C_{\omega}(t) d \omega-2 i \Omega C_{l}^{m}(t), \\
\\
\\
\dot{C}_{\omega}(t)=-i \alpha(\omega)\sum_{m=A, B}g_{m}^{*} e^{-i\left(\omega_{m}-\omega\right) t} \sum_{l=1}^{2} C_{l}^{m}(t),
\end{array}
\end{equation}
where $m\mathcal{}=A,B$ and $l\mathcal{}=1,2$ in the equation (A5). Next, we get $C_{\omega}(t)$ from the integration of the last part of equation (A5), and then substitute it into the remainder one. Therefor, we can obtain the following integro-differential equations
\begin{equation} \label{EB201}
\begin{aligned}
\frac{d C_{l}^{m}(t)}{d t}=&-\sum_{n=A, B} \int_{0}^{t} f_{m n}\left(t-t^{\prime}\right) \sum_{j=1}^{2} C_{j}^{n}\left(t^{\prime}\right) d t^{\prime} \\
&-2 i \Omega C_{l}^{m}(t),m=A,B,
\end{aligned}
\end{equation}
The kernel in equation (A6) is associated with the spectral density $J_{mn}(\omega ) $ of the reservoir, that is
\begin{equation} \label{EB201}
f_{m n}\left(t-t^{\prime}\right)=\int d \omega  J_{m n}(\omega) e^{i\left(\omega_{m}-\omega\right) t-i\left(\omega_{n}-\omega\right) t^{\prime}}.
\end{equation}
Let that the reservoir has the Lorentzian spectral density as
\begin{equation} \label{EB201}
J_{m n}(\omega)=\frac{1}{2 \pi} \frac{\gamma_{m n} \kappa ^{2}}{\left(\omega-\omega_{c}\right)^{2}+\kappa ^{2}},
\end{equation}
where $\gamma _{m n}=\frac{2 g_{m} g_{n}^{*}}{\kappa }$ is the relaxation rate of the excited state, and
\begin{equation} \label{EB201}
\gamma _{mm} =\gamma _{m} ,
\end{equation}
\begin{equation} \label{EB201}
\gamma_{m n}=\sqrt{\gamma_{m} \gamma _{n}} \theta, m \neq n,|\theta| \leq 1,
\end{equation}
where $\theta$ is defined as the SGI (spontaneously generated interference) parameter between the two decay channels $|A\rangle \to  |C\rangle$ and $|B\rangle \to  |C\rangle$ of each atom. The parameter $\theta$ depends on the angle between two dipole moments of the mentioned transitions.

Let that the two upper atomic states are degenerated and the atomic transitions are in resonant with the central frequency of the reservoir, i.e $\omega _{A} =\omega _{B} =\omega _{c}$, and $\gamma _{A}=\gamma _{B} =\gamma _{0} $ and $\gamma _{AB} =\gamma _{BA} =\gamma _{0} \theta$, $\gamma _{0}$ is the decay coefficient of the atomic excited state. The kernels in equation (A7) takes a simple form as
\begin{equation} \label{EB201}
\begin{aligned}
f_{A A}\left(t-t^{\prime}\right) &=f_{B B}\left(t-t^{\prime}\right)=f\left(t-t^{\prime}\right) \\
&=\int_{0}^{\infty} d \omega J(\omega) e^{i\left(\omega_{c}-\omega\right)\left(t-t^{\prime}\right)}, \\
f_{A B}\left(t-t^{\prime}\right) &=f_{B A}\left(t-t^{\prime}\right)=f^{\prime}\left(t-t^{\prime}\right) \\
&=\int_{0}^{\infty} d \omega J^{\prime}(\omega) e^{i\left(\omega_{c}-\omega\right)\left(t-t^{\prime}\right)},
\end{aligned}
\end{equation}
where $J^{\prime}(\omega)=\theta J(\omega )$.

Substituting equation (A8) into equation (A11), the following equations can be obtained
\begin{equation} \label{EB201}
f(t-t^{\prime})=\frac{\gamma _{0} \kappa }{2} e^{-\kappa  (t-t^{\prime})},
\end{equation}
\begin{equation} \label{EB201}
f^{\prime}(t-t^{\prime})=\frac{\gamma _{0} \theta \kappa }{2} e^{-\kappa  (t-t^{\prime})}.
\end{equation}

Here, $\gamma_{0}/\kappa<<1/2$ is the week coupling regime and $\gamma_{0}/\kappa>>1/2$ is the strong coupling regime \cite{Bellomo B}. Taking the Laplace transform from both sides of equation (A6), then we can get the following set of equations
\begin{equation} \label{EB201}
\begin{aligned}
 pC_{l}^{m}(p)-C_{l}^{m}(0)=&-\sum_{n=A, B} \mathcal{L}\left\{f_{m n}(t)\right\} \sum_{j=1}^{2} C_{j}^{n}(p) \\
&-2 i \Omega C_{l}^{m}(p), m=A, B,
\end{aligned}
\end{equation}
where $C_{l}^{m}(p)=\mathcal{L}\left\{C_{l}^{m}(t)\right\}=\int_{0}^{\infty} C_{l}^{m}(t) e^{-p t} d t$ is the Laplace transform of $C_{l}^{m}(t)$ and
\begin{equation} \label{EB201}
\mathcal{L}\{f(t)\}=\frac{\gamma _{0} \kappa }{2(\kappa +p)},
\end{equation}
\begin{equation} \label{EB201}
\mathcal{L}\left\{f^{\prime}(t)\right\}=\frac{\gamma _{0} \theta \kappa }{2(\kappa +p)}.
\end{equation}

Using equation (A14), we have the following equation
\begin{equation} \label{EB201}
(p+2 i \Omega) C_{l}^{m}(p)-C_{l}^{m}(0)=(p+2 i \Omega) C_{j}^{n}(p)-C_{j}^{n}(0),
\end{equation}
By considering this equation, equation (A14) can be written as
\begin{equation} \label{EB201}
\begin{aligned}
&(p+2 i \Omega) C_{l}^{m}(p)-C_{l}^{m}(0) \\
=&-\sum_{n= A,B}^{}\mathcal{L}\left\{f_{m n}(t)\right\}   \\
& \times\left(2 C_{l}^{n}(p)+\frac{1}{p+2 i \Omega}\sum_{j\ne l}^{2} \left(C_{j}^{n}(0)-C_{l}^{n}(0)\right)\right).
\end{aligned}
\end{equation}

After defining the new coefficient
\begin{equation} \label{EB201}
C_{l}^{\pm }(p)=C_{l}^{A} (p)\pm C_{l}^{B}(p),
\end{equation}
we can rewrite equation (A18) as
\begin{equation} \label{EB201}
\begin{aligned}
&(p+2 i \Omega) C_{l}^{\pm}(p)-C_{l}^{\pm}(0) \\
=&-\frac{\gamma _{0}(1 \pm \theta) \kappa }{2(\kappa +p)} \\
& \times\left(2 C_{l}^{\pm}(p)+\frac{1}{p+2 i \Omega}\sum_{j\ne l}^{2} \left(C_{j}^{\pm}(0)-C_{l}^{\pm}(0)\right)\right),
\end{aligned}
\end{equation}
i.e.
\begin{equation} \label{EB201}
\begin{aligned}
&(p+2 i \Omega) C_{1}^{\pm}(p)-C_{1}^{\pm}(0) \\
=&-\frac{\gamma _{0}(1 \pm \theta) \kappa }{2(\kappa +p)} \\
& \times\left(2 C_{1}^{\pm}(p)+\frac{1}{p+2 i \Omega}\left(C_{2}^{\pm}(0)-C_{1}^{\pm}(0)\right)\right),
\end{aligned}
\end{equation}
\begin{equation} \label{EB201}
\begin{aligned}
&(p+2 i \Omega) C_{2}^{\pm}(p)-C_{2}^{\pm}(0) \\
=&-\frac{\gamma _{0}(1 \pm \theta) \kappa }{2(\kappa +p)} \\
& \times\left(2 C_{2}^{\pm}(p)+\frac{1}{p+2 i \Omega}\left(C_{1}^{\pm}(0)-C_{2}^{\pm}(0)\right)\right).
\end{aligned}
\end{equation}
Taking inverse Laplace transform for equation (A20), giving the following equation
\begin{equation} \label{EB201}
\begin{aligned}
& C_{l}^{\pm}(t) \\
=& \mathcal{G_{\pm}}(t) C_{l}^{\pm}(0)-\frac{e^{-2 i \Omega t}-\mathcal{G_{\pm}}(t)}{2} \sum_{j \neq l}^{2}\left(C_{j}^{\pm}(0)-C_{l}^{\pm}(0)\right),
\end{aligned}
\end{equation}
i.e
\begin{equation} \label{EB201}
C_{1}^{\pm}(t)=\mathcal{G_{\pm}}(t) C_{1}^{\pm}(0)-\frac{e^{-2 i \Omega t}-\mathcal{G_{\pm}}(t)}{2}\left(C_{2}^{\pm}(0)-C_{1}^{\pm}(0)\right),
\end{equation}
\begin{equation} \label{EB201}
C_{2}^{\pm}(t)=\mathcal{G_{\pm}}(t) C_{2}^{\pm}(0)-\frac{e^{-2 i \Omega t}-\mathcal{G_{\pm}}(t)}{2}\left(C_{1}^{\pm}(0)-C_{2}^{\pm}(0)\right),
\end{equation}
where
\begin{equation} \label{EB201}
\begin{array}{l}
\mathcal{G_{\pm}}(t)=e^{-(\kappa +2 i \Omega) t / 2}
\\\times \left\{\cosh \left(\frac{D^{\pm} t}{2}\right)+\frac{\kappa -2 i \Omega}{D^{\pm}} \sinh \left(\frac{D^{\pm} t}{2}\right)\right\},
\end{array}
\end{equation}
and
\begin{equation} \label{EB201}
D^{\pm}=\sqrt{(\kappa +2 i \Omega)^{2}-4\left(2 i \Omega \kappa +\gamma _{0} \kappa (1 \pm \theta)\right)}.
\end{equation}
From equation (A19), we obtain the amplitudes
\begin{equation} \label{EB201}
C_{l}^{A} (t)=(C_{l}^{+}(t)+C_{l}^{-}(t))/2,
\end{equation}
\begin{equation} \label{EB201}
C_{l}^{B} (t)=(C_{l}^{+}(t)-C_{l}^{-}(t))/2.
\end{equation}

\


\begin{thebibliography}{200}
%%%%нд?  1-10
\bibitem{Pirandola S}Pirandola S, Eisert J, Weedbrook C, et al. Advances in quantum teleportation[J]. Nature photonics, 2015, 9(10): 641-652.
\bibitem{Bouwmeester D}Bouwmeester D, Pan J W, Mattle K, et al. Experimental quantum teleportation[J]. Nature, 1997, 390(6660): 575-579.
\bibitem{Yan Z H}Yan Z H, Qin J L, Qin Z Z, et al. Generation of non-classical states of light and their application in deterministic quantum teleportation[J]. Fundamental Research, 2021, 1(1): 43-49.
\bibitem{Cacciapuoti A S}Cacciapuoti A S, Caleffi M, Van Meter R, et al. When entanglement meets classical communications: Quantum teleportation for the quantum internet[J]. Institute of Electrical and Electronics Engineers Transactions on Communications, 2020, 68(6): 3808-3833.
\bibitem{Llewellyn D}Llewellyn D, Ding Y, Faruque I I, et al. Chip-to-chip quantum teleportation and multi-photon entanglement in silicon[J]. Nature Physics, 2020, 16(2): 148-153.
\bibitem{Lipka-Bartosik P}Lipka-Bartosik P, Skrzypczyk P. Catalytic quantum teleportation[J]. Physical Review Letters, 2021, 127(8): 080502.
\bibitem{C.H. Bennett}C.H. Bennett, G. Brassard, Quantum cryptography: Public key distribution and coin tossing[C]. Proc. International Conference on Computers, Systems and Signal Processing, 1984: 175-179.
\bibitem{Scarani V}Scarani V, Bechmann-Pasquinucci H, Cerf N J, et al. The security of practical quantum key distribution[J]. Reviews of Modern Physics, 2009, 81(3): 1301.
\bibitem{Renner R}Renner R. Security of quantum key distribution[J]. International Journal of Quantum Information, 2008, 6(01): 1-127.
\bibitem{Manzalini A}Manzalini A, Amoretti M. End-to-End Entanglement Generation Strategies: Capacity Bounds and Impact on Quantum Key Distribution[J]. Quantum Reports, 2022, 4(3): 251-263.
\bibitem{Fitzke E}Fitzke E, Bialowons L, Dolejsky T, et al. Scalable network for simultaneous pairwise quantum key distribution via entanglement-based time-bin coding[J]. PRX Quantum, 2022, 3(2): 020341.
\bibitem{Neumann S P}Neumann S P, Selimovic M, Bohmann M, et al. Experimental entanglement generation for quantum key distribution beyond 1 Gbit/s[J]. Quantum, 2022, 6: 822.
\bibitem{Nadlinger D P}Nadlinger D P, Drmota P, Nichol B C, et al. Experimental quantum key distribution certified by Bell's theorem[J]. Nature, 2022, 607(7920): 682-686.
\bibitem{DiVincenzo D P}DiVincenzo D P. Quantum computation[J]. Science, 1995, 270(5234): 255-261.
\bibitem{Nielsen M A}Nielsen M A, Chuang I L. Quantum computation and quantum information[J]. Cambridge University, 2001, 54(2): 60.
\bibitem{Liu B J}Liu B J, Song X K, Xue Z Y, et al. Plug-and-play approach to nonadiabatic geometric quantum gates[J]. Physical Review Letters, 2019, 123(10): 100501.
\bibitem{Shi Z C}Shi Z C, Wu H N, Shen L T, et al. Robust single-qubit gates by composite pulses in three-level systems[J]. Physical Review A, 2021, 103(5): 052612.
\bibitem{Kang Y H}Kang Y H, Song J, Xia Y. Error-resistant nonadiabatic binomial-code geometric quantum computation using reverse engineering[J]. Optics Letters, 2022, 47(16): 4099-4102.
\bibitem{Liu S}Liu S, Shen J H, Zheng R H, et al. Optimized nonadiabatic holonomic quantum computation based on Forster resonance in Rydberg atoms[J]. Frontiers of Physics, 2022, 17(2): 21502.
\bibitem{Mattle K}Mattle K, Weinfurter H, Kwiat P G, et al. Dense coding in experimental quantum communication[J]. Physical Review Letters, 1996, 76(25): 4656.
\bibitem{Guo H}Guo H, Liu N, Li Z, et al. Generation of continuous-variable high-dimensional entanglement with three degrees of freedom and multiplexing quantum dense coding[J]. Photonics Research, 2022, 10(12): 2828-2835.
\bibitem{Meher N}Meher N. Scheme for realizing quantum dense coding via entanglement swapping[J]. Journal of Physics B: Atomic, Molecular and Optical Physics, 2020, 53(6): 065502.
\bibitem{Liu B H}Guo Y, Liu B H, Li C F, et al. Advances in quantum dense coding[J]. Advanced Quantum Technologies, 2019, 2(5-6): 1900011.
\bibitem{Shaukat M I}Shaukat M I. Nonreciprocal quantum correlations and dense coding[J]. Physical Review A, 2022, 105(6): 062426.
\bibitem{Rijavec S}Rijavec S, Carlesso M, Bassi A, et al. Decoherence effects in non-classicality tests of gravity[J]. New Journal of Physics, 2021, 23(4): 043040.
\bibitem{Yu T}Yu T, Eberly J H. Sudden death of entanglement[J]. Science, 2009, 323(5914): 598-601.
\bibitem{Zou}Zou H M, Fang M F. Analytical solution and entanglement swapping of a double Jaynes-Cummings model in non-Markovian environments[J]. Quantum Information Processing, 2015, 14(7): 2673-2686.
%%%%11-20
\bibitem{Zou H M}Zou H M, Fang M F. Discord and entanglement in non-Markovian environments at finite temperatures[J]. Chinese Physics B, 2016, 25(9): 090302.
\bibitem{Mu Q}Mu Q, Lin P. Non-Markovian entanglement transfer to distant atoms in a coupled superconducting resonator[J]. Chinese Physics B, 2020, 29(6): 060304.
\bibitem{Faizi E}Behzadi N, Ahansaz B, Faizi E. Quantum coherence and entanglement preservation in Markovian and non-Markovian dynamics via additional qubits[J]. The European Physical Journal D, 2017, 71(11): 280.
\bibitem{Zhang Y J}Zhang Y J, Man Z X, Zou X B, et al. Dynamics of multipartite entanglement in the non-Markovian environments[J]. Journal of Physics B: Atomic, Molecular and Optical Physics, 2010, 43(4): 045502.
\bibitem{Flores M M}Flores M M, Galapon E A. Two qubit entanglement preservation through the addition of qubits[J]. Annals of Physics, 2015, 354: 21-30.
\bibitem{Mortezapour A}Nourmandipour A, Vafafard A, Mortezapour A, et al. Entanglement protection of classically driven qubits in a lossy cavity[J]. Scientific Reports, 2021, 11(1): 16259.
\bibitem{Wang Q}Wang Q, Liu R, Zou H M, et al. Entanglement dynamics of an open moving-biparticle system driven by classical-field[J]. Physica Scripta, 2022, 97(5): 055101.
\bibitem{Gholipour H}Mortezapour A, Nourmandipour A, Gholipour H. The effect of classical driving field on the spectrum of a qubit and entanglement swapping inside dissipative cavities[J]. Quantum Information Processing, 2020, 19(4): 136.
\bibitem{Zaffino R L}Maniscalco S, Francica F, Zaffino R L, et al. Protecting entanglement via the quantum Zeno effect[J]. Physical Review Letters, 2008, 100(9): 090503.
\bibitem{Wang X B}Wang X B, You J Q, Nori F. Quantum entanglement via two-qubit quantum Zeno dynamics[J]. Physical Review A, 2008, 77(6): 062339.
%%%%21-30
\bibitem{Liu R}Liu R, Zou H M, Yang J, et al. Entanglement witness and entropy uncertainty of an open quantum system under the Zeno effect[J]. Journal of the Optical Society of America B, 2021, 38(3): 662-669.
\bibitem{Kim Y S}Kim Y S, Lee J C, Kwon O, et al. Protecting entanglement from decoherence using weak measurement and quantum measurement reversal[J]. Nature Physics, 2012, 8(2): 117-120.
\bibitem{Wang S C}Wang S C, Yu Z W, Zou W J, et al. Protecting quantum states from decoherence of finite temperature using weak measurement[J]. Physical Review A, 2014, 89(2): 022318.
\bibitem{Liao X P}Liao X P, Fang M F, Fang J S, et al. Preserving entanglement and the fidelity of three-qubit quantum states undergoing decoherence using weak measurement[J]. Chinese Physics B, 2013, 23(2): 020304.
\bibitem{Zhang J}Zhang J, Wu R B, Li C W, et al. Protecting coherence and entanglement by quantum feedback controls[J]. Institute of Electrical and Electronics Engineers Transactions on Automatic Control, 2010, 55(3): 619-633.
\bibitem{Rafiee M}Rafiee M, Nourmandipour A, Mancini S. Universal feedback control of two-qubit entanglement[J]. Physical Review A, 2017, 96(1): 012340.
\bibitem{Liu Z}Liu Z, Kuang L, Hu K, et al. Deterministic creation and stabilization of entanglement in circuit QED by homodyne-mediated feedback control[J]. Physical Review A, 2010, 82(3): 032335.
\bibitem{Altintas F}Altintas F, Eryigit R. Dynamics of entanglement and Bell non-locality for two stochastic qubits with dipole-dipole interaction[J]. Journal of Physics A: Mathematical and Theoretical, 2010, 43(41): 415306.
\bibitem{Fasihi}Fasihi M A. Entanglement preservation in a system of two dipole-dipole interacting two-level atoms coupled with single mode cavity[J]. Physica Scripta, 2019, 94(8): 085104.
\bibitem{Ahansaz B}Ahansaz B, Behzadi N, Faizi E. Protection of entanglement for a two-qutrit V-type open system on the basis of system-reservoir bound states[J]. The European Physical Journal D, 2019, 73(3): 54.
%%%%31-40
\bibitem{Metwally N}Metwally N, Eleuch H, Obada A S. Sudden death and rebirth of entanglement for different dimensional systems driven by a classical random external field[J]. Laser Physics Letters, 2016, 13(10): 105206.
\bibitem{Xiao X}Xiao X, Li Y L. Protecting qutrit-qutrit entanglement by weak measurement and reversal[J]. The European Physical Journal D, 2013, 67(10): 204.
\bibitem{Wang M J}Wang M J, Xia Y J, Li Y D, et al. Protecting Qutrit-Qutrit entanglement under decoherence via weak measurement and measurement reversal[J]. International Journal of Theoretical Physics, 2020, 59(12): 3696-3704.
\bibitem{Li W J}Li W J, Zhao Y H, Leng Y. Protecting high-dimensional quantum entanglement from the amplitude-phase decoherence sources by weak measurement and reversal[J]. Laser Physics, 2019, 29(6): 065204.
\bibitem{Xia Y J}Wang M J, Xia Y J, Li Y D, et al. Protecting Two-qutrit Entanglement in Four Noise Channel Via Weak Measurement and Measurement Reversal[J]. International Journal of Theoretical Physics, 2021, 60(9): 3375-3386.
\bibitem{Bronn N T}Bronn N T, Magesan E, Masluk N A, et al. Reducing Spontaneous Emission in Circuit Quantum Electrodynamics by a Combined Readout/Filter Technique[J]. Applied Superconductivity Institute of Electrical and Electronics Engineers Transactions on, 2015, 25(5):1-10.
\bibitem{Vlastakis B}Vlastakis B, Petrenko A, Ofek N, et al. Characterizing entanglement of an artificial atom and a cavity cat state with Bell's inequality[J]. Nature communications, 2015, 6(1): 8970.
\bibitem{McKay D C}McKay D C, Naik R, Reinhold P, et al. High-contrast qubit interactions using multimode cavity QED[J]. Physical Review Letters, 2015, 114(8): 080501.
\bibitem{Hettrich M}Hettrich M, Ruster T, Kaufmann H, et al. Measurement of dipole matrix elements with a single trapped ion[J]. Physical Review Letters, 2015, 115(14): 143003.
\bibitem{Mok W K}Mok W K, You J B, Zhang W, et al. Control of spontaneous emission of qubits from weak to strong coupling[J]. Physical Review A, 2019, 99(5): 053847.
\bibitem{Takahashi H}Takahashi H, Kassa E, Christoforou C, et al. Strong coupling of a single ion to an optical cavity[J]. Physical Review Letters, 2020, 124(1): 013602.
\bibitem{Fink J M}Fink J M, Goppl M, Baur M, et al. Climbing the Jaynes-Cummings ladder and observing its nonlinearity in a cavity QED system[J]. Nature, 2008, 454(7202): 315-318.
\bibitem{Leek P J}Leek P J, Baur M, Fink J M, et al. Cavity quantum electrodynamics with separate photon storage and qubit readout modes[J]. Physical Review Letters, 2010, 104(10): 100504.
\bibitem{Stute A}Stute A, Casabone B, Schindler P, et al. Tunable ion-photon entanglement in an optical cavity[J]. Nature, 2012, 485(7399): 482-485.
\bibitem{Agarwal G S}Agarwal G S. Rotating-wave approximation and spontaneous emission[J]. Physical Review A, 1971, 4(5): 1778.
\bibitem{Behzadi N}Behzadi N, Faizi E, Heibati O. Quantum discord protection of a two-qutrit V-type atomic system from decoherence by partially collapsing measurements[J]. Quantum Information Processing, 2017, 16(10): 257.
\bibitem{Li C}Li C, Yang S, Song J, et al. Generation of long-living entanglement between two distant three-level atoms in non-Markovian environments[J]. Optics Express, 2017, 25(10): 10961-10971.
\bibitem{Li K}Li K, Dong C. Dynamic behaviors of coupled three-level atom system interacting with light field in cavity filled with Kerr-like medium[J]. Journal of Shanghai University, 2005, 9(4): 332-335.
\bibitem{Mandilara A}Mandilara A, Akulin V M. Cooperative behaviour of qutrits with dipole-dipole interactions[J]. Journal of Physics B: Atomic, Molecular and Optical Physics, 2007, 40(9): S95.
\bibitem{Peng J S}Peng J S, Li G X. Introduction to modern quantum optics[M]. World Scientific, 1998.
\bibitem{Mojaveri B}Mojaveri B, Dehghani A, Taghipour J. Control of entanglement, single excited-state population and memory-assisted entropic uncertainty of two qubits moving in a cavity by using a classical driving field[J]. The European Physical Journal Plus, 2022, 137(9): 1065.
\bibitem{Li Y}Li Y, Zhou J, Guo H. Effect of the dipole-dipole interaction for two atoms with different couplings in a non-Markovian environment[J]. Physical Review A, 2009, 79(1): 012309.
\bibitem{Ahmadi Z}Mojaveri B, Dehghani A, Ahmadi Z. A quantum correlated heat engine based on the parity-deformed Jaynes-Cummings model: Achieving the classical Carnot efficiency by a local classical field[J]. Physica Scripta, 2021, 96(11): 115102.
\bibitem{Fasihi M A}Mojaveri B, Dehghani A, Fasihi M A, et al. Ground state and thermal entanglement between two two-level atoms interacting with a nondegenerate parametric amplifier: Different sub-spaces[J]. International Journal of Modern Physics B, 2019, 33(06): 1950035.
\bibitem{Tavassoly M K}Faraji E, Baghshahi H R, Tavassoly M K. The influence of atomic dipole-dipole interaction on the dynamics of the population inversion and entanglement of two atoms interacting non-resonantly with two coupled modes field[J]. Modern Physics Letters B, 2017, 31(05): 1750038.
%%%%41-50
\bibitem{Nourmandipour A}Nourmandipour A, Tavassoly M K. Dynamics and protecting of entanglement in two-level systems interacting with a dissipative cavity: the Gardiner-Collett approach[J]. Journal of Physics B: Atomic, Molecular and Optical Physics, 2015, 48(16): 165502.
\bibitem{Bellomo B}Bellomo B, Franco R L, Compagno G. Non-Markovian effects on the dynamics of entanglement[J]. Physical Review Letters, 2007, 99(16): 160502.
\bibitem{Vidal G}Vidal G, Werner R F. Computable measure of entanglement[J]. Physical Review A, 2002, 65(3): 032314.
\bibitem{Wen Q}Wen Q. Formulas for partial entanglement entropy[J]. Physical Review Research, 2020, 2(2): 023170.
\bibitem{Wong G}Wong G, Klich I, Zayas L A P, et al. Entanglement temperature and entanglement entropy of excited states[J]. Journal of High Energy Physics, 2013, 2013(12): 20.
\bibitem{Li D}Li D, Liu M. Quantum Entanglement Death Problem Depict in Two Atomic Systems[J]. International Journal of Theoretical Physics, 2018, 57(5): 1265-1271.
\bibitem{Wootters W K}Wootters W K. Entanglement of formation of an arbitrary state of two qubits[J]. Physical Review Letters, 1998, 80(10): 2245.
\bibitem{Werner R F}Vidal G, Werner R F. Computable measure of entanglement[J]. Physical Review A, 2002, 65(3): 032314.
\bibitem{Calabrese P}Calabrese P, Cardy J, Tonni E. Entanglement negativity in quantum field theory[J]. Physical Review Letters, 2012, 109(13): 130502.



\end{thebibliography}
\end{document}